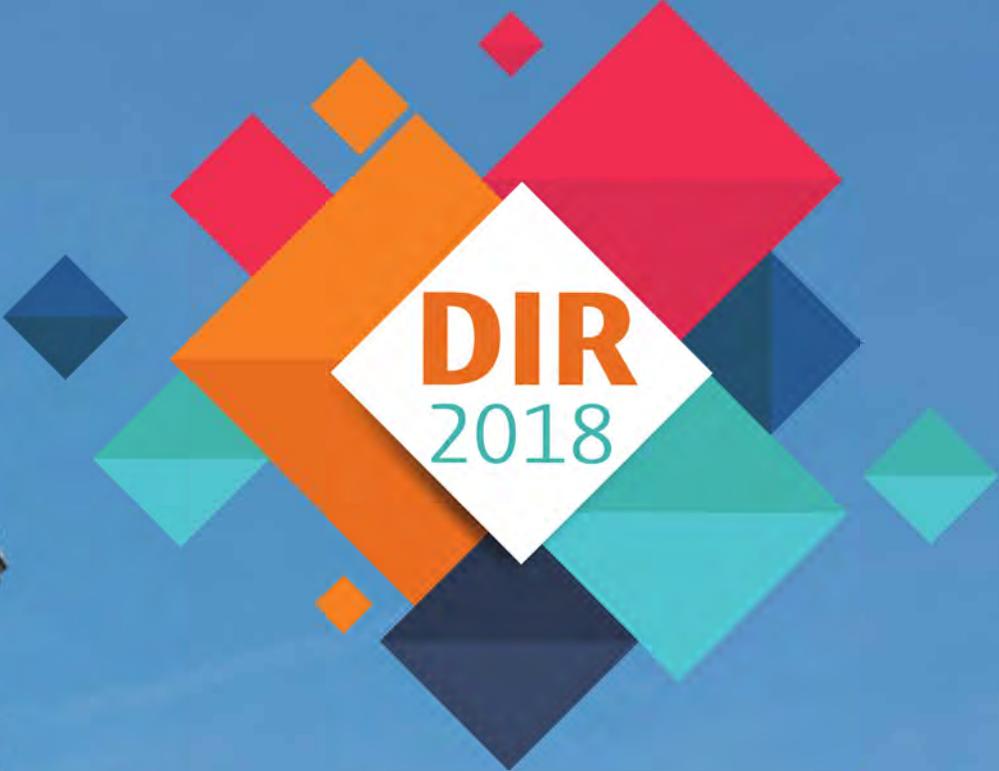

# DIR 2018

Proceedings of the 17th
Dutch-Belgian Information
Retrieval Workshop

23 November 2018
Leiden University

This volume contains the papers presented at DIR 2018: 17th Dutch-Belgian Information Retrieval Workshop (DIR) held on November 23, 2018 in Leiden. DIR aims to serve as an international platform (with a special focus on the Netherlands and Belgium) for exchange and discussions on research & applications in the field of information retrieval and related fields.

The committee accepted 4 short papers presenting novel work, 3 demo proposals, and 8 compressed contributions (summaries of papers recently published in international journals and conferences). Each submission was reviewed by at least 3 programme committee members. We thank the programme committee for their work.

From the accepted papers we compiled a programme that consisted of 2 keynotes, 5 oral presentations, 7 posters, and 3 demos.

Organising committee
- Suzan Verberne
- Roos van de Voordt
- Alex Brandsen
- Gineke Wiggers
- Hugo de Vos
- Wout Lamers
- Anne Dirkson
- Wessel Kraaij

Programme committee
- Toine Bogers
- Marieke van Erp
- David Graus
- Claudia Hauff
- Jiyin He
- Djoerd Hiemstra
- Jaap Kamps
- Udo Kruschwitz
- Florian Kunneman
- Martha Larson
- Edgar Meij
- Daan Odijk
- Roeland Ordelman
- Maya Sappelli
- Anne Schuth
- Dolf Trieschnigg
- Manos Tsagkias
- Arjen de Vries
- Wouter Weerkamp

# Table of contents

*Novel contributions*

*Compressed contributions*

# Lexical normalization of user-generated medical forum data

Anne Dirkson
Leiden University
a.r.dirkson@liacs.leidenuniv.nl

Suzan Verberne
Leiden University
s.verberne@liacs.leidenuniv.nl

Gerard van Oortmerssen
Leiden University
g.van.oortmerssen@liacs.leidenuniv.nl

Wessel Kraaij
Leiden University
w.kraaij@liacs.leidenuniv.nl

## ABSTRACT
In the medical domain, user-generated social media text is increasingly used as a valuable complementary knowledge source to scientific medical literature: it contains the unprompted experiences of the patient. Yet, lexical normalization of such data has not been addressed properly. This paper presents a sequential, unsupervised pipeline for automatic lexical normalization of domain-specific abbreviations and spelling mistakes. This pipeline led to an absolute reduction of out-of-vocabulary terms of 0.82% and 0.78% in two cancer-related forums. Our approach mainly targeted, and thus corrected, medical concepts. Consequently, our pipeline may significantly improve downstream IR tasks.

## 1 INTRODUCTION
In recent years, user-generated data from social media have been used extensively for medical text mining and information retrieval (IR) [4]. This user-generated data encapsulates a vast amount of knowledge, which has been used for a range of health-related applications, such as the tracking of public health trends [13] and the detection of adverse drug responses [12]. However, the extraction of this knowledge is complicated by non-standard and colloquial language use, typographical errors, phonetic substitutions, and misspellings [3, 11]. Social media text is generally noisy, and the complex medical domain aggravates this challenge [4]. The unique domain-specific terminology on forums cannot be captured by professional clinical terminologies because laypersons and healthcare professionals express health-related concepts differently [16].

Despite these challenges, normalization is one of the least explored topics in social media health language processing [4]. Medical lexical normalization methods, i.e. abbreviation expansion [6] and spelling correction [5, 10], have mostly been developed for clinical records or notes, as these also contain an abundance of domain-specific abbreviations and misspellings. However, social media text presents distinct challenges [4, 11] and cannot be tackled with these methods.

At the ACL W-NUT workshop in 2015, the best performing system for lexical normalization of generic social media combined rule-based and learning-based techniques [14]. Recently, Sarker [11] developed a modular pipeline that outperformed this system. His pipeline includes a customizable back-end module for domain-specific normalization, which employs spelling correction specifically for medical terms. However, it does not take into account that specialized forums often contain highly specific terms which may be excluded from the vocabulary. These terms are often essential for the task at hand (e.g. a novel drug name) and should thus not be 'corrected'. Additionally, Sarker [11] did not tackle domain-specific abbreviation expansion.

Thus, to further improve the quality of medical forum data, in this paper we will present two sequential domain-specific modules for lexical normalization of user-generated data, targeting abbreviations and spelling mistakes. The aim of this paper is two-fold. Firstly, we investigate to what extent these lexical normalization techniques can improve the quality of the patient forum text. Secondly, we apply these techniques to the second patient forum to test to what extent they are generalizable to other cancer-related medical forums.

## 2 DATA
### 2.1 Medical forum data
The first forum is a Facebook community, moderated by GIST Support International, an international patient forum for patients with Gastrointestinal Stromal Tumor (GIST). The data was collected in 2015 in collaboration with TNO. The second forum is the sub-reddit community on cancer, dating from 16/09/2009 until 02/07/2018.[1] It was scraped using the Pushshift Reddit API.[2] The data was collected in batches by looping over the timestamps in the data.

### 2.2 Abbreviations lexicon
Abbreviations were manually extracted from 500 randomly selected posts from the GIST data. This resulted in 47 unique abbreviations. For each abbreviation, two annotators firstly individually determined the correct expansion term, with an absolute agreement of 85.4%. Hereafter, they agreed on the correct form together. If ambiguous or context-dependent, the abbreviation was removed. For this reason, five abbreviations were removed.

---

[1]www.reddit.com/r/cancer
[2]https://github.com/pushshift/api

|              | # Tokens  | # Posts | Median length of post (IQR) |
|--------------|-----------|---------|-----------------------------|
| GIST forum   | 1,225,741 | 36,722  | 20 (35)                     |
| Reddit forum | 4,520,074 | 274,532 | 11 (18)                     |

Table 1: Raw data. The number of tokens and the median length of a post were calculated without punctuation.

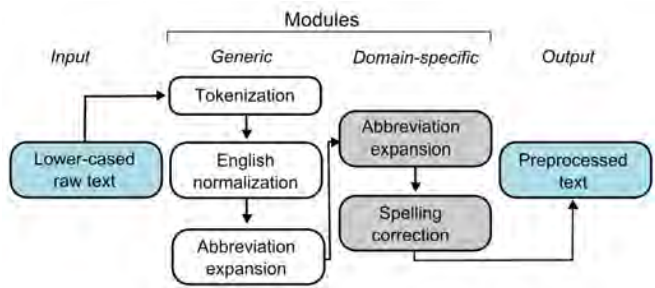

Figure 1: Sequential processing pipeline

### 2.3 Annotated data for spelling correction

The same 500 randomly selected posts were split into two sets of 250 posts: a tuning and a test set for detecting spelling mistakes. Each token was classified as a mistake (1) or not (0) by the first author. A second annotator checked if any of the mistakes were false positives. The first subset contained 34 unique non-word errors, equal to 0.39% of the tokens. Real-word errors, valid words used in the incorrect context, were not included. For the test set, these 34 mistakes and a tenfold of randomly selected correct words (340) with the same word length distribution were selected. The second subset contained 23 unique mistakes, equal to 0.31% of the tokens in the set. The tuning set consisted of these 23 mistakes combined with a tenfold of randomly selected correct words (230) with the same word length distribution. The tuning set was split in a stratified manner into 10 folds for cross-validation.

Combined, the two sets contained 55 unique mistakes: two mistakes occurred in both sets. The corrections of these mistakes were annotated individually by two annotators and then agreed on together. The absolute agreement was 89.0%. 8 mistakes were removed due to ambiguity (e.g. 'annonse' or 'gon'), resulting in 47 unique mistakes for evaluating the spelling correction algorithms.

## 3 METHODS
### 3.1 Preprocessing

To protect the privacy of users, in-text personal pronouns have been replaced as much as possible using a combination of the NLTK names corpus and part-of-speech tags (NNP and NNPS). Additionally, URLs and email addresses were replaced by the strings -url- and -email- using regular expressions. Furthermore, text was lower-cased and tokenized using NLTK. The first modules of the normalization pipeline of Sarker [11] were employed: converting British to American English spelling and the lexicon-based normalization of generic abbreviations. Some forum-specific additions were made: Gleevec (British variant: Glivec) was included in the first step and one generic abbreviation expansion that clashed with a domain-specific expansion was removed (i.e. 'temp' defined as *temperature* instead of *temporary*). Moreover, the Sarker dictionary was lower-cased and tokenized prior to preprocessing.

### 3.2 Abbreviation expansion

A simple lexicon lookup was used to expand the abbreviations in the data.

### 3.3 Spelling correction

We used the method by Sarker [11] (S1) as a baseline for spelling correction. His method combines normalized absolute Levenshtein distance (NAE) with Metaphone phonetic similarity and language model similarity. For the latter, distributed word representations (skip-gram word2vec) of three large Twitter datasets were used. It was compared with absolute Levenshtein distance (NAE), normalized as was done in S1, and relative Levenshtein distance (RE). Both were also explored with a penalty (-1) for differing first letters. Additionally, we investigated a version of Sarker's algorithm without language model similarity (S2).

We manually constructed a decision process, inspired by the work by Beeksma [1], for detecting spelling mistakes. The decision process makes use of a token's frequency in the corpus, and the similarity with possible replacements. The underlying idea is that if a word is common within the domain-specific language or there is no similar enough candidate available, it is unlikely to be a mistake.

To ensure generalisability, we opted for an unsupervised, data-driven method that does not rely on the construction of a specialized vocabulary. For measuring similarity and correcting terms, the generic CELEX lexicon [2] was combined with all corpus tokens surpassing the frequency threshold. The latter are considered only after the CELEX terms and in order of frequency (from high to low). Of the candidates with the highest similarity score, the first is selected.

To optimize the decision process, a 10-fold cross validation grid search of the maximum relative corpus frequency [1E-6, 2.5E-6, 5E-6, 1E-5, 2E-5, 4E-5] and maximum relative edit distance (0.15 to 0.25 with 0.01 increments) was conducted with the tuning set. The choice of grid was based on previous work by Walasek [15] and Beeksma [1]. The loss function used to tune the parameters was the $F_{0.5}$ score, which places more weight on precision than the $F_1$ score. We believe it is more important to not alter correct terms, than to retrieve incorrect ones.

### 3.4 Evaluating data quality

The percentage of out-of-vocabulary (OOV) terms is used as an estimation of the quality of the data: less OOV-terms and thus more in-vocabulary (IV) terms reflects cleaner data. To calculate the number of OOV terms, a merged vocabulary was created by combining the standard English lexicon CELEX [2], the NCI Dictionary of Cancer Terms [7], the generic and commercial drug names from

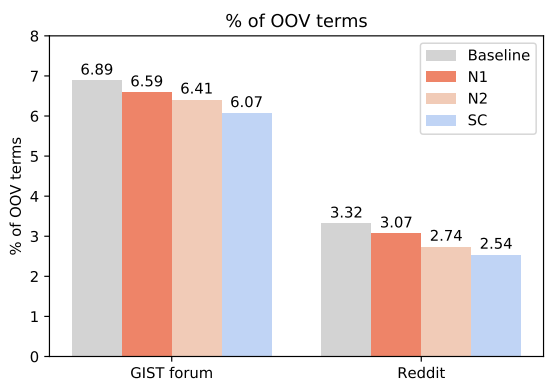

**Figure 2: Number of OOV-terms with sequential modules. N1: Generic abbreviation expansion [11]. N2: Domain-specific abbreviation expansion. SC: Spelling correction.**

the RxNorm [8], the ADR lexicon used by Nikfarjam et al. [9] and our abbreviation expansions. [3]

## 4 RESULTS
### 4.1 Abbreviation expansion

The baseline % of OOV-terms was higher for the GIST data (6.9%) than the Reddit data (3.3%). The most effective reduction of OOV-terms in both forums was achieved by combined generic and domain-specific abbreviation expansion (N1+N2) (see Fig 2). This was slightly more effective in the Reddit (-0.58%) than the GIST data (-0.47%) (see Fig. 2).

The additional domain-specific abbreviation expansion replaced 4747 terms distributed over 3756 posts (18.7% of the data) in the GIST forum and 18688 terms in 16479 posts (6.0% of the data) in the Reddit forum. The associated OOV-term reduction was 0.18% and 0.33% resp. The replacements did not appear concentrated in a small number of posts in either forum: respectively 81.3% and 88.9% of the posts with replacements had a single replacement.

31 of the 36 abbreviations found in the GIST forum were also present in the Reddit forum, indicating that these abbreviations are to some extent generalizable between cancer-related forums. The abbreviations that were not present in the cancer sub-reddit were: hpfs (*high power fields*), vit (*vitamin*), gf (*girlfriend*), mg/d (*mg/day*) and til (*until*). There was also large overlap (80%) between the ten most common abbreviation expansions in the forums. For the Reddit forum, chemotherapy (69.9%) was by far the most common expansion. Although a common treatment for many cancers, it is an uncommon treatment for GIST, which explains the relative low frequency (5.7%) for the GIST forum.

### 4.2 Spelling correction

*Detecting spelling mistakes.* The grid search resulted in a max. corpus frequency of 5E-06 and a max. similarity score of 0.19 (see Table 2). This combination attained the maximum $F_{0.5}$ score for all

[3]available at urlhttps://github.com/AnneDirkson/lex_normalization

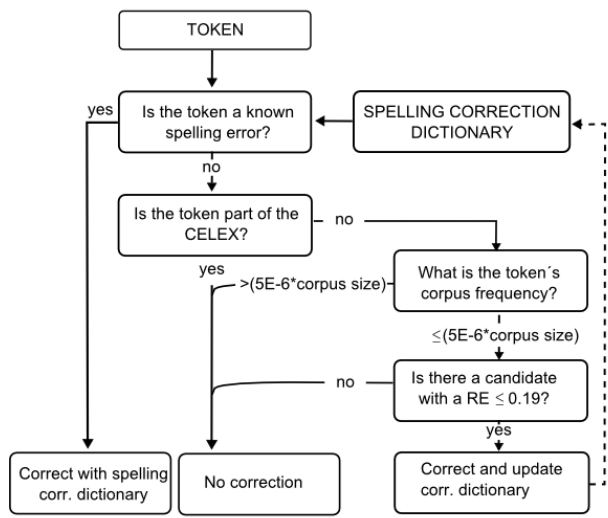

**Figure 3: Decision process for spelling corrections. RE: Relative Edit Distance**

|  |  | Recall | Precision | $F_1$ | $F_{0.5}$ | AUC |
|---|---|---|---|---|---|---|
| **CELEX** | Test | 0.94 | 0.51 | 0.66 | 0.56 | 0.92 |
| **Decision** | Validation | 0.62 | 0.76 | 0.67 | 0.72 | 0.80 |
| **process** | Test | 0.38 | 1.0 | 0.55 | 0.75 | 0.69 |

**Table 2: Detection of spelling mistakes. The average of a 10-fold CV was taken for the validation set.**

| False negatives | abdomin | oncogolgist | metastisis | thanx |
|---|---|---|---|---|
| True positives | oncolgy | clenical | metastized | surgry |

**Table 3: Examples of false negatives (i.e. missed mistakes) and true positives (i.e. found mistakes) found in the test set using mistake detection with the decision process**

|  | NAE | NAE+P | RE | RE+P | S1 | S2 |
|---|---|---|---|---|---|---|
| Accuracy | 59.6% | 59.6% | **66.0%** | **66.0%** | 23.4% | 19.1% |
| Duration (s) | 6.09 | 7.29 | 3.84 | 4.07 | 257.00 | 237.42 |

**Table 4: Spelling correction. NAE: normalized absolute edit distance. +P: with first-letter penalty. RE: relative edit distance. S1: Sarker's algorithm S2: S1 without language model similarity. Duration was measured over an average of 5 runs.**

folds. Despite a low recall on the test set (0.38), the precision was 1. Thus, although mistakes may be missed, no correct terms are falsely marked as errors. Unfortunately, this does mean that some common mistakes, like oncogolgist, are missed (see Table 3).

*Comparing spelling correction algorithms.* Relative edit distance (RE) was the most accurate spelling correction algorithm (66.0%) (see Table 4). The first-letter penalty did not improve the accuracy.

| Mistake | gleevac | opnion | sutant | kontrol |
|---|---|---|---|---|
| Correction | gleevec | opinion | sutent | control |
| NAE | **gleevec** | option | mutant | **control** |
| NAE+P | **gleevec** | option | **sutent** | kowtow |
| RE | **gleevec** | **opinion** | mutant | **control** |
| RE+P | **gleevec** | **opinion** | **sutent** | kestrel |
| S1 | colonic | option | mutant | contr |
| S2 | gleeful | option | mutant | controls |

Table 5: Examples of spelling correction results. NAE: normalized absolute edit distance. +P: with first-letter penalty. RE: relative edit distance. S1: Sarker's algorithm. S2: S1 without the language model.

Since the corrections of four mistakes did not occur in the vocabulary, the upper bound of accuracy was 91.5%. Interestingly, the two versions of Sarker's method (S1 and S2) managed to correct only 23.4% and 19.1% of the mistakes respectively. This showcases the limitations of using generic social media normalization techniques in the medical domain.

*Evaluating the spelling correction module.* In the GIST data, 3367 mistakes were replaced with 2601 unique terms. The mistakes often concern important medical terms. The ten most frequent corrections were: gleevec (17x), oncologist (13x), diagnosed (10x), positive (8x), stivarga (8x), imatinib (8x), metastasized (7x), regorafenib (7x) and tumors (7x). Gleevec, stivarga, imatinib and regorafenib are cancer medications.

In the Reddit forum, 5238 mistakes were replaced with 4161 unique terms, of which the most prevalent were: metastasized (10x), treatment (10x), diagnosed (10x), adenocarcinoma (10x), symptoms (9x), immunotherapy (9x), lymphoma (8x), patients (8x), dexamethasone (8x) and cannabinoids (8x). Thus, our module appears to effectively target medical terms.

The reduction in OOV-terms was higher for the GIST (0.34%) than for the Reddit forum (0.20%) (See Fig. 2). Furthermore, our method only targets infrequent spelling mistakes: in both forums, all corrected spelling mistakes occurred only once.

## 5 DISCUSSION

For domain-specific abbreviation expansion and sequential spelling correction, the combined reduction in OOV-terms was 0.59% and 0.54% for the GIST and Reddit forum resp. Although this reduction may seem minor, our approach mainly targets medical concepts, which are highly relevant for downstream tasks such as named entity extraction. The pipeline appears generalizable for cancer-related forums: it resulted in comparable reductions in OOV-terms for both forums.

The generic lexical normalization pipeline by Sarker [11] does not appear to suffice for normalizing health-related user-generated text. We identified 36 additional domain-specific abbreviations in our data that were not corrected in their method. Moreover, our analysis revealed that their spelling correction algorithm performed poorly compared to both relative and absolute Levenshtein distance. One must note, however, that the test set excluded real-word errors, slang and ambiguous errors.

Our study has a number of limitations. Firstly, the use of OOV-terms as a proxy for quality of the data relies heavily on the vocabulary that is chosen and, moreover, does not allow for differentiation between correct and incorrect substitution of words. In the future, we will instead opt for extrinsic performance measures to investigate the utility of our approach. Secondly, our data-driven spelling correction could lead to the 'correction' of spelling mistakes with other spelling mistakes. This possibility cannot be excluded entirely, but is countered by sorting the corpus tokens on frequency. A larger tuning set could perhaps improve the thresholding.

## 6 CONCLUSION

Our sequential unsupervised pipeline can improve the quality of text data from medical forum posts. Future work will explore the impact of our pipeline on task performance using established benchmark data from diverse medical forums.

## REFERENCES

[1] M. Beeksma. 2017. *Computer: how long have I got left?* Master's thesis. Radboud University, Nijmegen, the Netherlands.
[2] G. Burnage, R.H Baayen, R. Piepenbrock, and H. van Rijn. 1990. CELEX: A Guide for Users. (1990).
[3] E. Clark and K. Araki. 2011. Text Normalization in Social Media: Progress, Problems and Applications for a Pre-processing System of Casual English. *Procedia Soc Behav Sci* 27 (2011), 2–11. https://doi.org/10.1016/j.sbspro.2011.10.577
[4] G. Gonzalez-Hernandez, A. Sarker, K. O'Connor, and G. Savova. 2017. Capturing the Patient's Perspective: a Review of Advances in Natural Language Processing of Health-Related Text. *Yearbook of medical informatics* (2017), 214–217. https://doi.org/10.15265/IY-2017-029
[5] K.H. Lai, M. Topaz, F.R. Goss, and L. Zhou. 2015. Automated misspelling detection and correction in clinical free-text records. (2015). https://doi.org/10.1016/j.jbi.2015.04.008
[6] D.L. Mowery, B.R. South, L. Christensen, J. Leng, L.M. Peltonen, S. Salanterä, H. Suominen, D. Martinez, S. Velupillai, N. Elhadad, G. Savova, S.r Pradhan, and W. W. Chapman. 2016. Normalizing acronyms and abbreviations to aid patient understanding of clinical texts: ShARe/CLEF eHealth Challenge 2013, Task 2. *Journal of Biomedical Semantics* (2016). https://doi.org/10.1186/s13326-016-0084-y
[7] National Cancer Institute. [n. d.]. NCI Dictionary of Cancer Terms. https://www.cancer.gov/publications/dictionaries/cancer-terms
[8] National Library of Medicine (US). [n. d.]. RxNorm. https://www.nlm.nih.gov/research/umls/rxnorm/
[9] A. Nikfarjam, A. Sarker, K. O'Connor, R. Ginn, and G. Gonzalez. 2015. Pharmacovigilance from social media: mining adverse drug reaction mentions using sequence labeling with word embedding cluster features. *Journal of the American Medical Informatics Association : JAMIA* 22, 3 (2015), 671–81. https://doi.org/10.1093/jamia/ocu041
[10] J. Patrick, M. Sabbagh, S. Jain, and H. Zheng. 2010. Spelling correction in clinical notes with emphasis on first suggestion accuracy. *2nd Workshop on Building and Evaluating Resources for Biomedical Text Mining* (2010), 2–8.
[11] A. Sarker. 2017. A customizable pipeline for social media text normalization. *Social Network Analysis and Mining* 7, 45 (2017). https://doi.org/10.1007/s13278-017-0464-z
[12] A. Sarker, R. Ginn, A. Nikfarjam, K. O'Connor, K. Smith, S. Jayaraman, T. Upadhaya, and G. Gonzalez. 2015. Utilizing social media data for pharmacovigilance: A review. *Journal of Biomedical Informatics* 54 (2015), 202–212. https://doi.org/10.1016/J.JBI.2015.02.004
[13] A. Sarker, K. O'Connor, R. Ginn, M. Scotch, K. Smith, D. Malone, and G. Gonzalez. 2016. Social Media Mining for Toxicovigilance: Automatic Monitoring of Prescription Medication Abuse from Twitter. *Drug Safety* 39, 3 (2016), 231–240. https://doi.org/10.1007/s40264-015-0379-4
[14] D. Supranovich and V. Patsepnia. 2015. IHS_RD: Lexical Normalization for English Tweets. In *Proceedings of the ACL 2015 Workshop on Noisy User-generated Text*. 78–81.
[15] N. Walasek. 2016. *Medical Entity Extraction on Dutch forum data in the absence of labeled training data.* Master's thesis. Radboud University, Nijmegen, the Netherlands.
[16] Q. Zeng and T. Tse. 2006. Exploring and developing consuming health vocabulary. *J Am Med Inform Assoc* 13, 1 (2006), 24–29. https://doi.org/10.1197/jamia.M1761.A

# Exploration of Intrinsic Relevance Judgments by Legal Professionals in Information Retrieval Systems

Gineke Wiggers*
eLaw - Center for Law and Digital Technologies
Leiden University
Leiden, The Netherlands
g.wiggers@law.leidenuniv.nl

Suzan Verberne
Leiden Institute for Advanced Computer Science
Leiden University
Leiden, The Netherlands
s.verberne@liacs.leidenuniv.nl

Gerrit-Jan Zwenne
eLaw - Center for Law and Digital Technologies
Leiden University
Leiden, The Netherlands
g.j.zwenne@law.leidenuniv.nl

## ABSTRACT
This paper addresses relevance in legal information retrieval (IR). We study the factors that influence the perception of relevance of search results for users of Dutch legal IR systems. These factors can be used to improve the ranking of search results, so that legal professionals will find the information they need faster. The relevance factors are identified by a user questionnaire in which we showed users of a legal IR system a query and two search results. The user had to choose which of the two results he/she would like to see ranked higher for the query and was asked to provide a reasoning for their choice. The search results were chosen in the manner of a vignette, to test two potentially relevant factors. The questionnaire had eleven pairs of search results spread over two queries. 43 legal professionals participated in our study. This method has proven to make the options different enough for users to seriously consider both and give indications of their relevance assessment process. The tested and reported factors were mostly part of the algorithmic, topical and cognitive relevance sphere. Consensus on these factors means that developers of legal IR systems can incorporate these factors into their ranking algorithms.

## CCS CONCEPTS

• **Information systems** → **Information retrieval**; **Specialized information retrieval**; **Relevance assessment**;

## KEYWORDS
Legal information retrieval, Expert search, Relevance, User study

**ACM Reference Format:**
Gineke Wiggers, Suzan Verberne, and Gerrit-Jan Zwenne. 2018. Exploration of Intrinsic Relevance Judgments by Legal Professionals in Information Retrieval Systems. In *Proceedings of Dutch-Belgian Information Retrieval Workshop (DIR2018).* ACM, New York, NY, USA, 4 pages.

---

*Gineke Wiggers is PhD candidate at Leiden University and business analyst at Legal Intelligence.

## 1 INTRODUCTION

Relevance, in the broadest sense, is a term used to describe "Connection with the subject or point at issue; relation to the matter in hand." [1] In everyday language, it is used to describe the effectiveness of information in a given context. [10, p. 203] In information retrieval, the theory of relevance has several dimensions, including algorithmic relevance, topical relevance, cognitive relevance, situational relevance, and, in particular for legal information retrieval (IR), bibliographic relevance.[12]

Literature [7] suggests that users of (legal) IR systems have implicit criteria for the relevance/value judgments about documents presented to them. This is supported by anecdotal evidence from employees of Legal Intelligence, one of two large legal content integration and IR systems in the Netherlands. Users of the Legal Intelligence system have reported a preference of documents with certain characteristics over others, for example a preference for recent case law over older case law, case law from higher courts over case law from lower courts; sources which are considered authoritative (government publications) over blogs or news items, well-known authors over lesser-know authors, and/or the official version (case law or law) over reprints.

Previous studies addressing relevance criteria conducted user observation studies with a thinking-aloud protocol or interviews, or a combination of both. [3][11][8][5] These studies are time consuming, and therefore difficult to conduct with legal professionals whose hours are expensive. This research proposes a method to make explicit which factors or criteria users intrinsically consider when assessing a search result in legal IR, using a focused questionnaire with pairwise comparisons that could be completed by a legal expert in 12 minutes. The outcome of this study will allow for exploration of these factors and their occurrence across subgroups of users. It is conducted with users of the Legal Intelligence[1] system.[2] The study addresses the following research questions:

(1) Is a questionnaire with forced choice a suitable method to explore factors that influence the perception of relevance of users in legal IR systems?
(2) What factors influence the perception of relevance of users of Dutch legal IR systems?

The answers to these questions will show whether this method is suitable for exploring these factors of relevance. If suitable, the found factors will allow the improvement of precision in legal IR

---

[1]www.legalintelligence.com
[2]For the importance of testing with real users of the IR system, see Park [7, p. 322]

systems - which are often focused on recall - and indicate what future research should focus on.

The contributions of this paper compared to previous work are: (1) we propose a method for eliciting the implicit relevance criteria that users of search systems have; (2) we conducted a user study with professional users of a legal IR system; (3) we show that there is consensus among the users about the criteria they use for judging the relevance of legal documents; (4) we confirm previous exploratory work and the anecdotal evidence given by users of a Dutch legal IR system.

## 2 BACKGROUND

Relevance criteria have been investigated before in the context of web search. Already in 1998, Rieh and Belkin [8] addressed the user's perception of quality and authority as relevance factors. In 2006, Savolainen and Kari [11] found in an exploratory study that specificity, topicality, familiarity, and variety were the four most mentioned criteria in user-formulated relevance judgments, but there was a high number of individual criteria mentioned by the participants.

This work is done in the context of the theory on spheres of relevance as described by Saracevic [9] and Cosijn and Ingwersen [4], and applied to the legal domain by Van Opijnen and Santos [12]. The spheres of relevance that play a role in legal IR are algorithmic relevance, topical relevance, cognitive relevance, situational relevance and bibliographic relevance.

This research attempts to explore the factors that influence the perception of relevance as proposed by Barry [3]. Compared to the work of Barry, we investigate context in an expert domain (legal IR) as opposed to open-domain web search. Methodologically, we use a forced-decision questionnaire with pre-set relevance criteria that were hidden to the participants, as opposed to open interviews used by Barry. Because of this method, our study focuses on algorithmic, topical and cognitive relevance rather than the situational relevance of the user. The choice to use actual users, rather than domain experts was influenced by Park [7, p. 322]. The chosen method, with examples that vignette-like differ on certain characteristics but are the comparable on other characteristics, was inspired by work by Atzmüller and Steiner [2].

## 3 METHODS

The questionnaire consisted of three parts. The first part covered general questions regarding the legal field the respondent is active in, his/her function profile, and his/her level of expertise.

For each of the next two parts of the questionnaire, the respondents are shown an example search query. Because the judgment of relevance of results is in large part influenced by the perceived information need of the query [9, p. 340] (the cognitive relevance), respondents are first asked to indicate what information need they think the user is trying to fulfill by issuing this query.

To mitigate the effect of this situational relevance the questionnaire uses two example queries rather than the respondent's actual information needs and tasks.[3] It is expected that with example queries respondents will indicate factors related to algorithmic, topical and cognitive relevance, which will allow for a general analysis of the user group as a whole[6, p. 37].

### 3.1 Relevance factors

The factors chosen for the questions are from the algorithmic, topical and cognitive relevance spheres. In the setup of the questionnaire each possible relevance factor occurs in at least two pairs of search results, three if the factor has three levels. The tested factors were:

- Recency [3, p. 156]: it has been suggested that recent case law is more relevant than older case law (< 2 years; 2 - 10 years; > 10 years old), but it can also be related to the specific period the case played in [12, p. 80];
- Legal hierarchy/importance [12, p. 68]: case law from higher courts carries more weight than case law from lower courts (supreme court; courts of appeal; courts of first instance);
- Annotated[4]: annotated case law (providing context for the case) is more relevant than case law that is not annotated;
- Source authority[5] [3, p. 156]: sources that are considered authoritative are preferred over other sources (government documents, leading publications; mid-range publications; blogs);
- Authority author[6] [3, p. 155-156]: documents written by well-known authors are considered more authoritative than other documents;
- Bibliographical relevance [12, p. 71]: the official version (case law or law) is more relevant than reprints;
- Title relevance: results with search term in the title or summary are considered more relevant than results with the search term not in the title/summary (the visibility of algorithmic relevance for respondents);
- Document type[7]: document types that pertain to the perceived information need are considered more relevant than other document types (depending on perceived information need expressed in the query as interpreted by respondent).

The respondents were not informed which relevance factors were tested in the paired results. The factors were not mentioned explicitly in any stage of the questionnaire.

### 3.2 Selection of stimuli

We manually selected the two example queries from the query logs of the Legal Intelligence search engine. Both queries are broadly recognizable, so that all respondents will have an understanding of the information need the user is trying to fulfill, and the (type of) documents that can fulfill this need. The queries serve as context for the respondent, but the tested factors are query independent, with the exception of the factor whether the document type matches the information need as perceived by the respondent. To exclude query bias, all respondents are shown the same two queries.

---

[3]In contrast to, for example, Barry [3] who used information needs from users.

[4]As mentioned by users to Legal Intelligence employees.
[5]Also described as Source quality.
[6]Also described as Relationship with author and Source reputation/visibility
[7]Van Opijnen and Santos [12, p. 68] mentions the large diversity in document types in legal IR.

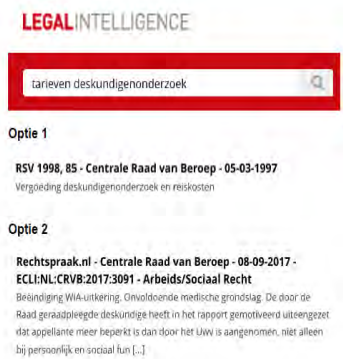

**Figure 1: A screenshot of the questionnaire. The example query is shown in the query field on top and the two search results (choices) are listed as 'optie 1' and 'optie 2' below.**

The respondents are shown the query along with two related search results, shown as images from actual search results as they are displayed in the legal IR system. The interface of the pairwise choices is illustrated in Figure 1. The search results were chosen in the manner of a vignette study where all results include at least two of the relevance factors that are mentioned in the literature as relevance factors for Legal IR (see section 3.1).

Respondents are asked to give a relative relevance judgment by indicating which of the two results he/she would like to see ranked higher than the other. We chose relative scoring because research by Saracevic shows that relative scoring leads to more consistent results across respondents of different backgrounds than individual document scorings [9, p. 341].

Where authoritative sources or authors are tested, it was attempted to show sources and authors that are so generally known that respondents from other legal fields will likely recognize these names from their legal education, or can estimate it by the academic title of the author. It is assumed that the other tested factors of the relevance judgment, such as whether a case is annotated, are valid for all legal fields.

Though it is expected that the factors mentioned by the respondents reflect these factors, respondents are given a free text field to give their own motivation for their choice. Research has shown that the primitive/intuitive definition of relevance prevails when respondents are confronted with questions regarding relevance judgment [10, p. 203]. For that reason, no formal definition of relevance was given in the questionnaire. In the introduction of the questionnaire some examples of factors were given[8]. To avoid leading the respondents, and to encourage respondents to consider both results from their own perspective, these examples were not repeated alongside the questions.

It is likely that a relevance judgment based on title and summary (as in this research) differs from the relevance judgment upon reading the entire document. [9, p. 340] Because this research focuses on perceived relevance for the purpose of ranking in IR systems - which document a user is more likely to open - the research focuses on the perceived relevance of titles/summaries as shown in the IR system.

A preliminary pilot questionnaire suggested that the target audience prefers a questionnaire that can be completed in under 12 minutes. Because of this, the number of queries is limited to two and it was not possible to show all possible combinations of factors. Each participant saw eleven pairs of search results spread over the two example queries. This did not hinder the research, since the purpose of this research is to understand the factors that influence the perception of relevance, rather than generalizing findings or making predictions.

### 3.3 Participants

All users of the Legal Intelligence IR system were able to fill out the questionnaire. The questionnaire was made available online, so that respondents could fill it out at a moment convenient for them, to ensure maximum response. It was distributed to the national government and large law firms through their information specialists, and to all other users by a newsletter and a LinkedIn post. The survey was brought to the attention of acquaintances who work in the legal field via email. We aimed for 50 responses, distributed over the different affiliation types, law area specialisms, and roles.

## 4 RESULTS

43 people completed the survey, leading to a total of (43*11=) 473 choices made. The participants came from a range of areas of legal expertise, function types, organization types and years of work experience.

Though the considerations given by the respondents often differed from the factors for which the corresponding examples were chosen, it seems that the factors behind the two options make the choices different enough for respondents to seriously consider both choices. On average, respondents are split 31:12 over the choices. The highest agreement reached for a choice was a division of 40:3[9], and the lowest agreement was 20:23.

Respondents could give a free text explanation for each of the choices. Often, these contained one or more factors, or a statement indicating the respondent had no preference. In 90 instances, there was no (clear) explanation.

We manually grouped the answers according to the factors that we defined in Section 3.1. Because respondents were not asked to describe the weight of the factors in the outcome of their choice or whether it was the determining factor, the frequency of the factor does not indicate its importance, only how often it was mentioned. All mentioned factors are listed in Table 1.

The found factors confirm the tested factors. In addition, it appears that the level of depth or detail of a document[11] and the law area of the document (as determined through the title, source or author) are considered when determining which document respondents wish to see ranked higher. Though the method focused on algorithmic, topical and cognitive relevance, users also mentioned the usability of the document as a factor, and the length of the document[12]. These factors appear to be more related to the situational relevance, and were therefore not part of the tested factors. It is interesting to see that there appears to be a collective cognitive relevance, what Van Opijnen and Santos [12] call domain relevance, in legal IR consisting of factors like source authority, legal hierarchy and whether a document is annotated.

---

[8]Translated the examples read: 'This could be because the title or summary seems more relevant, the result comes from an authoritative source, the publication date of the document, or because it is a document type where you expect to find the answer to the query.'

[9]With no clear common factor of the three respondents who chose option 2.
[10]Also related to background/experience as described by Barry [3, p. 156]
[11]Described by Barry [3, p. 156] as depth/scope

**Table 1: Relevance factors sorted by number of mentions in the free text field. An asterisk (*) indicates that the factor was not one of the tested factors (listed in Section 3.1) but added by participants**

| Factor | Number of times mentioned |
| --- | --- |
| Title relevance | 154 |
| Document type | 67 |
| Recency | 59 |
| Level of depth* | 58 |
| Legal hierarchy | 44 |
| Law area (topic)[10] | 31 |
| Authority/credibility (total) | 31 |
|     Source authority | 15 |
|     Authority author | 9 |
| Usability | 15 |
| Bibliographical relevance | 12 |
| Annotated | 7 |
| Length of the document | 2 |
| No preference | 28 |

We have tested whether function type, law area, the amount of work experience and the type of organization a respondent works for has an impact on his/her responses. This appears to be limited.

## 5 DISCUSSION

It is interesting to note that document type is the second most named consideration for the respondent's relevance choices. This suggests that when legal professionals are searching for something, they know what type of document they are likely to find the information in. Similarly, the level of depth respondents are looking for (fourth most reported argument) influences what document types they open. Legal IR systems often do not reflect this in their results list, focusing on algorithmic relevance and gathering results from all document types in one list. They rely on filtering options to guide users to the information they are looking for.

The most reported consideration, whether the word is in the title or summary of the result, shows that simple changes in the user interface might already improve the perception of the quality of the ranking for users, without actually changing the ranking itself. By showing snippets (where the section of the document where the query terms are found is shown) on the search results page, rather than publisher curated summaries as is currently the case in the system and examples used for this research, users will be able to see the query terms in context, and better understand the relevance of the document.

---

[12]Described by Barry [3, p. 156] as time constraints

Considering that that the majority of respondents in each group generally chose the same option as the majority in the other groups, it seems that a number of these factors (the collective cognitive or domain relevance factors) can be used to improve the ranking of legal IR systems on a general level. Incorporating the lessons learned from this research could be a first step to enhance the user experience, while further research is conducted into further incorporating cognitive and potentially situational relevance into legal IR systems.

## 6 CONCLUSIONS

RQ 1. Is a questionnaire with forced choice a suitable method to explore factors that influence the perception of relevance of user in legal IR systems?

We found that a vignette style questionnaire with forced choice appears to be a suitable method to explore factors that influence the perception of relevance of user in legal IR systems. Compared to a user observation study or interviews, a forced choice-questionnaire costs less time for the participants and allows us to control the stimuli and investigate the factors we are interested in.

RQ 2. What factors influence the perception of relevance of users of Dutch legal IR systems?

The factors that influence the perception of relevance of users of Dutch legal IR systems are title relevance, document type, recency, level of depth, legal hierarchy, law area (topic), authority/credibility, usability, bibliographical relevance, annotated and length of the document. The factors found confirm the conclusions of the 25-year old user study by Barry [3] and the anecdotal evidence given by Legal Intelligence users.

In the near future we will use the outcome of this research to validate improvement to ranking algorithms in legal IR systems.

## ACKNOWLEDGMENTS
The authors wish to thank the employees of Legal Intelligence for their cooperation in this research.

## REFERENCES
[1] [n. d.]. Oxford English Dictionary. ([n. d.]).
[2] C. Atzmüller and P.M. Steiner. 2010. Experimental Vignette Studies in Survey Research. *Methodology* 6, 3 (2010), 128–138.
[3] C.L. Barry. 1994. User-Defined Relevance Criteria: An Exploratory Study. *Journal of the American Society for Information Science* 45, 3 (1994), 149–159.
[4] E. Cosijn and P. Ingwersen. 2000. Dimensions of relevance. *Information Processing and Management* 36 (2000), 533–550.
[5] P. Ingwersen and K. Järvelin. 2005. Information retrieval in context: IRiX. *Acm sigir forum* 39, 2 (2005), 31–39.
[6] D.A. Kemp. 1973. Relevance, Pertinence and Information System Development. *Information Storage and Retrieval* 10 (1973), 37–47.
[7] T.K. Park. 1993. The Nature of Relevance in Information Retrieval: an Empirical Study. *Library Quarterly* 63, 3 (1993), 318–351.
[8] S. Y. Rieh and N. J. Belkin. 1998. Understanding judgment of information quality and cognitive authority in the WWW. In *Proceedings of the 61st annual meeting of the American society for information science.* 279–289.
[9] T. Saracevic. 1975. Relevance: A Review of and Framework for the Thinking on the Notion in Information Science. *Journal of the American Society for Information Science* 1975 (1975), 321–343.
[10] T. Saracevic. 1996. Relevance Reconsidered, Information Science: Integration in perspectives. In *Proceedings of the Second Conference on Conceptions of Library and Information Science.* 201–218.
[11] R. Savolainen and J. Kari. 2006. User-defined relevance criteria in web searching. *Journal of Documentation* 62, 6 (2006), 685–707.
[12] M. van Opijnen and C. Santos. 2017. On the concept of relevance in legal information retrieval. *Artificial Intelligence and Law* 25 (2017), 65–87.

# JudaicaLink; A Domain-Specific Knowledge Base for Jewish Studies

Maral Dadvar and Kai Eckert
Web-based Information Systems and Services (WISS)
Stuttgart Media University, Germany
{dadvar , eckert}@hdm-stuttgart.de

## ABSTRACT

JudaicaLink is a novel resource which provides a knowledge base of Jewish literature, culture and history. It is based on multilingual domain-specific information from encyclopedia and general-purpose knowledge bases such as DBPedia. The main goal of JudaicaLink is the contextualization of metadata of digital collections, i.e., entity resolution within and linking of metadata to improve access to digital resources and to provide a richer context to the user. Many resources for contextualization, particularly specialized resources for the given domain, are only available in unstructured form. General-purpose resources such as DBpedia are hard to use due to their sheer size while only a very small subset of the data is actually relevant. Therefore, JudaicaLink aims at integrating relevant subsets of various data sources to function as a single hub for the contextualization process. JudaicaLink is freely available on the Web as Linked Open Data. In this paper, we briefly explain how JudaicaLink is built, how it can be accessed by users, as well as its architecture, technical implementation, and applications. We hope that through this paper we reach out to the Dutch-Belgium information retrieval community and get to know other potential relevant sources which can be integrated and further enrich our knowledge base.

## CCS CONCEPTS

• **Information systems** → **Digital libraries and archives** • **Information systems** → **Semantic web description languages** • Computing methodologies → Semantic networks

## KEYWORDS

Knowledge Extraction; Information Retrieval; Linked Open Data

## 1 INTRODUCTION

A knowledge base is a collection of knowledge about variety of entities and it contains facts explaining those entities [11]. Besides being used for applications such as question answering [7], semantic search [12], visualization [9], and machine translation [8], knowledge bases also play an important role in information integration. Some knowledge bases are specific to a certain domain such as occupations and job activities [5], others are general such as DBPedia [1] and Yago [2] which are huge sources of structured knowledge extracted from Wikipedia and other sources.

In this paper, we introduce JudaicaLink[3], a new knowledge base specific to Jewish culture, history and studies. With JudaicaLink, we build a domain-specific knowledge base by extracting structured, multilingual knowledge from different sources. The main application of JudaicaLink so far is the contextualization of metadata, i.e., entity resolution within and linking of metadata to improve resource access and to provide richer context to the user. The task of contextualization consists of two steps; first, to identify entities unambiguously by means of stable URIs, e.g., a corresponding DBpedia resource, and the second to find as much information (e.g., descriptions, links to related entities) as possible about the identified entity, usually by following links (such as owl:sameAs) to other data sources and this way by obtaining further URIs suitable for identification.

Many useful data sources exist that can be used for contextualization in the domain of Jewish studies, e.g., domain-specific online encyclopedias like the YIVO Encyclopedia of Jews in Eastern Europe[4]. In contrast to Wikipedia, they describe all entities in depth from the domain perspective, i.e., with respect to Jewish history, which makes them more useful for our task. On the other hand they lack the broad coverage of Wikipedia and the structured data access via Linked Open Data representations such as DBpedia or Yago. Additionally, there are highly relevant data sources such as the Integrated

---

[1] http://wiki.dbpedia.org/
[2] http://yago-knowledge.org/
[3] http://www.judaicalink.org/
[4] http://yivoencyclopedia.org/

Authority File (GND) of the German National Library[5] providing mainly unique identifier, but also brief additional contextual information, usually of a very high quality. An unexpected drawback of these knowledge bases, however, is their sheer size. Setting up DBpedia or the GND for a local contextualization process is not a trivial task and requires considerable technical resources, despite the fact that only a very small portion of these knowledge bases are relevant to the domain of Jewish studies.

In particular, there are three main problems that need to be dealt with; First, unstructured data sources like online encyclopedias need to be made available as structured data with stable URIs. Second, relevant subsets of general-purpose knowledge bases like DBpedia have to be identified to fill the gaps between the specialized resources and to provide further context. And last, all data sources have to be integrated and interlinked.

JudaicaLink is RDF-based [3] and part of the Linked Open Data cloud. It includes information about persons, geographic places, subjects and occupations. At the time of this writing it contains 43,690 persons and 23,068 concepts. All data is available via our public SPARQL endpoint and as data dumps.

## 2 CONSTRUCTION OF JUDAICALINK

In this section we will describe the data sources which are integrated into JudaicaLink. We will explain the pros and cons of encyclopedias and general-purpose knowledge bases as data sources. Moreover, the infrastructure of the knowledge base, the data extraction process and representation are briefly explained.

### 2.1 Sources

Reference works such as encyclopedias and glossaries function as guides to specific scholarly domains. Therefore encyclopedias with a focus on Jewish studies were one of the sources of information in our knowledge base. The following encyclopedias have been so far integrated into JudaicaLink. What all these encyclopedias have in common is that they did not exist in a structured data format before. By using customized web scrapers, we extracted structured data and our required information from the article pages, e.g., the title, the article text, link relations to other articles.

**Encyclopedia of Russian Jewry**. Encyclopedia of Russian Jewry[6] provides an Internet version of the encyclopedia, which is published in Moscow since 1994, giving a comprehensive, objective picture of the life and activity of the Jews of Russia, the Soviet Union and the CIS. The encyclopedia is structurally divided into three parts: 1. biographical information, 2. local history of the Jewish community in pre-revolutionary Russia, the Soviet Union and the CIS, and 3. thematic information on concepts related to Jewish civilization, the contribution of the Jews of Russia in various fields of activity, various Jewish social, scientific, cultural organizations, etc. The originally published volumes contain more than 10,000 biographies and more than 10,000 place names. The electronic version contains corrections and additions in the form of new articles, all in all 20,434 concepts.

**Yivo Encyclopedia**. The YIVO Encyclopedia of Jews in Eastern Europe[3], courtesy of the YIVO Institute of Jewish Research, provides articles concerned with the history and culture of Jews in Eastern Europe from the beginnings of their settlement in the region to the present. This online source makes accurate, reliable, scholarly information about East European Jewish life accessible to everyone. The dataset contains 2,374 concepts.

**Das Jüdische Hamburg**. Das Jüdische Hamburg[7] is an encyclopedia containing articles in German by notable scholars about persons, locations and events of the history of Jewish communities in Hamburg. Das Jüdische Hamburg is a free online resource based on the book "Das Jüdische Hamburg - Ein historisches Nachschlagewerk" [6]. It was published in 2006 on the occasion of the 40th anniversary of the Institute for the History of the German Jews[8]. It is a comparatively small dataset of 260 concepts.

**Biographisches Handbuch der Rabbiner.** The Biographisches Handbuch der Rabbiner is an online encyclopedia provided by the Salomon L. Steinheim-Institute for German-Jewish history at the University of Duisburg-Essen, edited by Michael Brocke and Julius Carlebach. The goal of this encyclopedia is to be a complete directory of all rabbis who lived and worked in or originated from German-speaking areas since the age of enlightenment. The encyclopedia consists of two parts [1, 2]. This dataset contains more than 2,900 persons.

For extraction of the encyclopedias' contents we have made use of Coffeescript, Javascript and Python modules. To this

---

[5] http://dnb.de/
[6] http://rujen.ru/
[7] http://dasjuedischehamburg.de/
[8] Institut für die Geschichte der deutschen Juden, IGdJ

end, regular expression based methods were used for extraction of information such as birth date, death date, birth location, death location and occupation. Here we should emphasize on rich interlinking between the datasets. There are also knowledge bases which contain a vast variety of information including facts related to Jewish culture. Therefore we also used these sources to extract a focused knowledge graph of concepts for the domain of Jewish studies:

**DBpedia**. DBpedia is a large-scale source of structured and multilingual knowledge extracted from Wikipedia. This knowledge base contains over 400 million facts that describe 3.7 million things [10]. We follow several approaches to extract relevant concepts from DBpedia: our main focus so far was on identifying prominent Jewish persons from different fields of activities. By identifying categories used to describe Jewish persons, we generated a list of these categories and searched for further persons. For each person, we extracted the name in all available languages, as well as links to other data sources. Typical categories include occupations, like "Rabbi".

As occupations are often available in other sources as well, we created occupation ontology, combining labels and other information from various sources. The DBpedia dataset contains currently 5,294 persons with 35 distinct occupations.

**GND**. The Integrated Authority File (GND) of the German National Library is an authority file that contains among other identifiers for persons. Unlike DBPedia with its many categories, Jewish persons are not distinguished by any means in GND. Strategies to find relevant entries include the exploitation of publication data where the relevance can be determined via the publication. Occupations can also be used, but to a much smaller extent than in DBpedia, as DBpedia often contains specific categories for "Jewish authors", for instance, where GND only contains "author" as occupation. We also considered geographic information where available, for example for persons from Israel. For every person the name, occupation and identifiers were extracted. In the resulted RDF file the persons and their corresponding attributes were mapped to JudaicaLink ontology. This dataset includes 4,029 persons and 303 occupations.

To extract the domain-specific graphs from the mentioned knowledge bases we used python code modules. All the extraction and data generation codes are available open source on our GitHub repository[9]. In the resulting RDF files the persons and their corresponding attributes were mapped to JudaicaLink ontology.

## 2.2 Infrastructure

JudaicaLink provides the datasets in N3 (Notation3) and its subset formats, Turtle (Terse RDF Triple Language, TTL) . This format facilitates the usage and integration of JudaicaLink in triple stores and Semantic Web software such as Apache Jena. The main JudaicaLink website is driven by the static site generator Hugo. We use the metadata of the web pages (Hugo frontmatter) to control the data publication process which is fully automated. On every push to the master branch, GitHub triggers an update script on our server that pulls the latest changes, rebuilds the website using Hugo and updates the data in the triple store according to the page metadata. This way we ensure that the dataset descriptions on the web site, the data dumps and the data loaded in JudaicaLink are always consistent. Every dataset corresponds to a name graph that can later on be accessed in the triple store. Datasets may consist of more than one data file since they might have been further expanded over time or may content different data components. Users can download JudaicaLink datasets from the webpage of JudaicaLink. The datasets can also be browsed as Linked Open Data using Pubby (with DM2E extensions) as Web Frontend [9]. Furthermore, a public SPARQL endpoint [10] is available.

## 2.3 Ontology

The classes and properties used in JudaicaLink ontology[11] are created on the fly based on the information that we encounter and need to be represented. However, we are consistent on the usage of the properties and the coined URI's are stable and unique. When a piece of information described in an encyclopedia is extracted, we assign the class 'Concept'. We use NLP techniques to analysis the concepts in order to identify whether they are a person. When identified as such, the class 'Person' is assigned to them and further properties are added. Every property of a Concept can be also used for a Person.

---

[9] https://github.com/wisslab/judaicalink-loader/
[10] http://data.judaicalink.org/sparql
[11] https://tinyurl.com/yal5wa2b

## 3 APPLICATIONS OF JUDAICALINK

**FID Judaica**. The FID Judaica project[12] aims to create an expert information service for the domain of Jewish studies. This service creates a portal as a central platform for scientific information and among other purposes it aims to contextualize its extensive digital Judaica collections. JudaicaLink is used as the source by which the metadata of these collections are enriched. So far 39.8% of the library digital collection has been contextualized using JudaicaLink. We manually evaluated 10% percent of the contextualized records. The contextualization accuracy was 0.93.

**Automatic Identification of Jewish Studies Titles**. Another application of JudaicaLink is for automatic classification of Jewish studies titles [4]. There is a large number of titles in library collections which are not classified and indexed. In order to automatically identify and extract the Jewish titles from such collections, a Natural Language Processing based classification tool was developed. JudaicaLink was used to determine the relevance of the identified author name, which was used as one of the features, based on the available information in the knowledge base. The overall accuracy of the classification model was 89%.

## 4 CONCLUSION AND FUTURE WORK

In this paper we presented JudaicaLink, a knowledge base for Jewish literature and culture, which merges domain-specific information from different sources into one coherent entity. We described sources, the extraction process, and the applications of JudaicaLink. As future work, we would like to extend this project along different directions. We plan to extend our information extraction by textual analysis. The textual sources, such as the abstracts and definitions, contain additional information which needs to be first identified and extracted and then turned into triples. The ontology used in our knowledge base will be improved and more specific as we progress and come across new facts and entities. JudaicaLink is an ever growing source of information and adding new relevant resources is a continual goal towards making JudaicaLink a rich and comprehensive reference. We hope that through this paper we reach out to the Dutch-Belgium information retrieval community and while introducing this novel knowledge base we also get to know other potential relevant sources which can be integrated and further enrich our knowledge base.

## ACKNOWLEDGMENTS

MD is supported by Fachinformationsdienst Jüdische Studien project funded by German Research Foundation (DFG, Projektnummer 286004564).

## REFERENCES

[1] Brocke, M. and Carlebach, J. 2004. *Biographisches Handbuch der Rabbiner; Teil 1 Die Rabbiner der Emanzipationszeit in den deutschen, böhmischen und großpolnischen Ländern 1781–1871*. K. G. Saur Verlag GmbH.

[2] Brocke, M. and Carlebach, J. 2004. *Biographisches Handbuch der Rabbiner; Teil 2 Die Rabbiner im Deutschen Reich 1871–1945*. K. G. Saur Verlag GmbH.

[3] Cyganiak, R., Wood, D. and Lanthaler, M. 2017. RDF 1.1 concepts and abstract syntax. *W3C Recommendation*. (2017).

[4] Dadvar, M., Heuberger, R., Sasse, A. and Eckert, K. 2017. Automatic NLP-based Classification of Jewish Studies Titles. *In 41st European Library Automation Group Conference* (Athens, 2017).

[5] Gassen, J.B., Faralli, S., Ponzetto, S.P. and Mendling, J. 2016. Who-does-what: A knowledge base of people's occupations and job activities. *CEUR Workshop Proceedings* (2016), Paper--65.

[6] Heinsohn, K. 2006. *Das juedische Hamburg: ein historisches Nachschlagewerk*. Wallstein Verlag.

[7] Kiyota, Y., Kurohashi, S. and Kido, F. 2002. Dialog navigator: A question answering system based on large text knowledge base. *Proceedings of the 19th international conference on Computational linguistics-Volume 1* (2002), 1–7.

[8] Knight, K. and Luk, S.K. 1994. Building a Large-Scale Knowledge Base for Machine Translation. *AAAI* (1994), 773–778.

[9] Kraker, P., Kittel, C. and Enkhbayar, A. 2016. Open Knowledge Maps: Creating a Visual Interface to the World's Scientific Knowledge Based on Natural Language Processing. *027.7 Journal for Library Culture*. 4, 2 (2016), 98–103. DOI:https://doi.org/10.12685/027.7-4-2-157.

[10] Lehmann, J., Isele, R. and Jakob, M. 2014. DBpedia–A large-scale, multilingual knowledge base extracted from Wikipedia. *Semantic Web*. 00, 2 (2014), 1–29. DOI:https://doi.org/10.3233/SW-140134.

[11] Rebele, T., Suchanek, F., Hoffart, J., Biega, J., Kuzey, E. and Weikum, G. 2016. YAGO: A multilingual knowledge base from wikipedia, wordnet, and geonames. *Lecture Notes in Computer Science (including subseries Lecture Notes in Artificial Intelligence and Lecture Notes in Bioinformatics)* (2016), 177–185.

[12] Tran, T., Cimiano, P., Rudolph, S. and Studer, R. 2007. Ontology-Based Interpretation of Keywords for Semantic Search. *The Semantic Web*. (2007), 523–536. DOI:https://doi.org/10.1007/978-3-540-76298-0_38.

---

[12] https://ub.uni-frankfurt.de/projekte/juedische_studien

# Recommending Users:
# Whom to Follow on Federated Social Networks

Jan Trienes
University of Twente
Enschede, Netherlands
j.trienes@student.utwente.nl

Andrés Torres Cano
University of Twente
Enschede, Netherlands
a.f.torrescano@student.utwente.nl

Djoerd Hiemstra
University of Twente
Enschede, Netherlands
d.hiemstra@utwente.nl

## ABSTRACT
To foster an active and engaged community, social networks employ recommendation algorithms that filter large amounts of contents and provide a user with personalized views of the network. Popular social networks such as Facebook and Twitter generate follow recommendations by listing profiles a user may be interested to connect with. Federated social networks aim to resolve issues associated with the popular social networks – such as large-scale user-surveillance and the miss-use of user data to manipulate elections – by decentralizing authority and promoting privacy. Due to their recent emergence, recommender systems do not exist for federated social networks, yet. To make these networks more attractive and promote community building, we investigate how recommendation algorithms can be applied to decentralized social networks. We present an offline and online evaluation of two recommendation strategies: a collaborative filtering recommender based on BM25 and a topology-based recommender using personalized PageRank. Our experiments on a large unbiased sample of the federated social network Mastodon shows that collaborative filtering approaches outperform a topology-based approach, whereas both approaches significantly outperform a random recommender. A subsequent live user experiment on Mastodon using balanced interleaving shows that the collaborative filtering recommender performs on par with the topology-based recommender.

## 1 INTRODUCTION
Evergrowing concerns about user-privacy, censorship and central authority in popular social media have motivated both the development of federated social networks such as Mastodon and Diaspora [2, 10], as well as research in academia [1, 11]. These networks aim to promote user control by decentralizing authority and relying on open-source software and open standards. At the time of this writing, Mastodon has over 1 million users and 3500 instances which demonstrates the increasing acceptance of distributed social networks. As with traditional social media, one key success factor of such a network is an active and engaged community.

As a community grows, overwhelming amounts of content make it increasingly difficult for a user to find interesting topics and other users to interact with. For that reason, popular platforms such as Twitter, LinkedIn and Facebook introduce recommender systems that set out to solve a particular recommendation task. One prominent example is the "Who to Follow" service by Twitter [4]. Due to their recent emergence, those recommender systems do not exist for federated social networks, yet. However, they are needed to make distributed social media attractive to large user groups as well as competitive to centralized networks. At the same time, recommender systems will contribute to develop, grow and sustain an active community.

To make federated social networks more attractive and feature complete, we implement and evaluate a topology-based user recommender based on personalized PageRank [9], a commonly used algorithm for link-prediction in social networks. We compare this method against collaborative filtering based on link intersections [5] and a random link predictor baseline [7]. The experiments are carried out on Mastodon, a federated social network for which user relations do not require reciprocation, and the network forms a directed graph. We expect that the method and results are transferable to any other federated social network with similar characteristics.

We evaluate the systems in an offline and online scenario. For the offline evaluation, we collect an unbiased sample of the Mastodon user graph. This sample is created by performing a Metropolis-Hastings Random Walk (MHRW) adapted for directed graphs [12, 13]. The collected data contains about 25% of the entire userbase of Mastodon. We then evaluate the recommender systems according to standard performance metrics used in ranked retrieval systems, and deploy the two best performing methods to an online setting. Both algorithms generate a list of personalized recommendations for 19 Mastodon users participating in the online trial and performance is measured with the balanced interleaving approach [6].

This paper is structured as follows. Section 2 explains how data are collected for the offline experiments and discusses the recommendation algorithms and their evaluation. In Section 3 we present and discuss experimental results. Section 4 concludes this paper and provides directions for future work.

## 2 DATASET AND METHODS
### 2.1 Recommendation Algorithms
The user recommendation problem for social networks can be formalized as follows. Given a graph $G = (V, E)$ where $V$ and $E$ are vertices and edges, we seek to predict an interaction between a user $u \in V$ and $v \in V$ denoted by edge $(u, v)$. In networks such as Mastodon and Twitter, a user interaction does not require reciprocation. Thus, the graph is directed. We consider two broad approaches to generate recommendations: (1) collaborative filtering-based recommendation and (2) topology-based recommendation.

With respect to the collaborative filtering, we use an approach inspired by [5]. Each user $u \in V$ is represented by a profile and recommendations are generated based on the similarity of profiles. We distinguish between the three best performing strategies in [5]:

*following*($u$)   The set of user ID's $u$ follows
*followers*($u$)   The set of user ID's that follow $u$
*combined*($u$)   The combined set of following and follower ID's

We consider these profiles as documents to be indexed in a general purpose search engine. In order to generate recommendations for a user, the corresponding profile is extracted first. Afterwards, the retrieval system is queried with the profile and it ranks the indexed documents by their relevance to the query. Each ID in the user profile is a token of the query. If a query consists of more than 10,000 tokens, we create a random subset of 10,000 tokens. Unlike Hannon et al. [5], we use BM25 instead of TF-IDF to estimate the relevance score of each document and set parameters to common defaults ($k_1 = 1.2$, $b = 0.75$) [8]. The final recommendation list contains the top-$k$ documents with highest retrieval score.

The collaborative filtering recommendations are compared to topology-based recommendations. Several methods have been proposed in literature which make use of link-based ranking algorithms such as HITS, PageRank and SALSA. Due to the novelty of generating recommendations for federated social networks, we restrict our experiments to the personalized PageRank algorithm [9] whose efficient computation is well-understood and which is used in the Twitter recommender system [4]. We apply the personalized PageRank for a seed node which is the user we want to generate recommendations for. After convergence, the list of user recommendations is constructed by taking the top-$k$ nodes with highest PageRank. Following [7], we set the damping factor $\lambda = 0.85$.

## 2.2 Data Collection

Acquiring the complete graph of a social network is always infeasible due to API limits and time constraints [13]. An additional concern arises in a distributed social network. As data is not stored at a central authority, there is no single API that provides access to all parts of the network. Instead, data is scattered around different sub-networks. Both issues are addressed within this section.

To overcome the time constraint, we apply the Metropolis-Hastings Random Walk (MHRW) to acquire an unbiased sample that is still representative of the complete graph. MHRW is a Markov-Chain Monte Carlo algorithm that can be used to obtain node samples with a uniform probability distribution [13]. As the MHRW is only applicable to undirected graphs, we apply a generalization that considers all directed edges as bidirectional edges [12]. We do not consider graph sampling methods such as Random Walk and Breadth-First Sampling as it has been shown that these methods yield samples biased towards high degree nodes [3].

Due to the fact that a distributed social network has no central API, one has to query the API of each individual sub-network referred to as *instance*. In case of Mastodon, there are two public endpoints to acquire incoming and outgoing links: /following and /followers[1]. Whenever the MHRW visits an unexplored node, followers and followings of that node are fetched and stored in a document-oriented database. This database is also used as a cache: if the random walk transitions to a node which it has already visited, we use the cached result rather than querying the API again. During the data collection, we apply fair crawling policies. Only instances that allow crawling as defined by the robots.txt are considered. Furthermore, concurrent requests are throttled such that no more than 10 requests per second are issued (a rate which we believe any web server can sustain).

---
[1]The following API URL pattern applies to any Mastodon instance: https://<instance>/users/<user>/<endpoint>.json

Table 1: Statistics of crawled graphs. The initial crawl at $t_1$ and the newer crawl of the same users at $t_2$.

| Graph | $|V|$ | $|V^*|$ | $|E|$ | Assort. | Deg. | NCC | SCC |
|---|---|---|---|---|---|---|---|
| $t_1$ | 253,822 | 3437 | 754,037 | -0.015 | 5.94 | 0.31 | 0.175 |
| $t_2$ | 255,638 | 3383 | 754,667 | -0.016 | 5.9 | 0.31 | 0.173 |

## 2.3 Dataset Statistics

Table 1 summarizes the properties of the collected graph. The initial graph ($t_1$) has been crawled from the 16/05/18 until 17/05/18. The MHRW was executed for 5500 iterations. During the crawl, 138 instances were disregarded either because of their robots.txt or because they were no longer available. In order to acquire a newer version of that graph ($t_2$), we visited the same users five days later and recorded new relationships. The number of visited users in $t_2$ is slightly lower than in $t_1$, as some profiles were deleted or their instances became unavailable. The updated graph is used as the ground-truth when evaluating our recommender systems.

It can be observed that the Network Average Clustering Coefficient (NCC) and the fraction of nodes in the largest Strongly Connected Component (SCC) is almost equal for the two given graphs. Furthermore, the graph is mildly disassortative. It is important to mention that although the total number of nodes found $|V|$ is high (253,000), accounting for about 25% of the total Mastodon users, the number of visited nodes $|V^*|$ is much smaller (about 3400). Incoming and outgoing edges are only known for visited nodes.

## 2.4 Evaluation

The algorithms presented in Section 2.1 are evaluated in two phases: an offline evaluation and an online evaluation. For the offline evaluation we measure precision at rank $k$ (p@k), Mean Average Precision (MAP) and success at rank $k$ (s@k), which are popular metrics for the evaluation of ranked retrieval systems [8]. The newer graph at time $t_2$ serves as the ground-truth, whereas the graph at time $t_1$ can be seen as the training graph. In information retrieval terms, the generated list of recommendations are the retrieved documents and the list of users a target user follows at time $t_2$ are the relevant documents. Significance is tested using a two-tailed paired t-test. We denote improvements with ▲($p < 0.01$), deteriorations with ▼($p < 0.01$), and no significance by °.

During the offline evaluation, all systems generate a list of 100 recommendations based on the training graph at time $t_1$. This list is then compared with the actual links added to the graph in between time $t_1$ and $t_2$ (see Section 2.3). In case of the collected dataset, 329 of 3437 visited users started to follow another individual, and thus added a link to the graph. Only for this set of users, recommendations are generated and evaluated.

Table 2: Experimental results of offline evaluation. Significance for model in line $i > 1$ is tested against line $i - 1$.

| ID | System | MAP | s@1 | s@5 | s@10 |
|----|--------|-----|-----|-----|------|
| R1 | Random | 0.001 | 0.000 | 0.000 | 0.055 |
| R2 | Profile (following) | **0.019**▲ | **0.033**▲ | 0.085▲ | 0.152▲ |
| R3 | Profile (followers) | **0.019**° | 0.030° | 0.100° | 0.167° |
| R4 | Profile (combined) | 0.018° | **0.033**° | **0.106**° | **0.173**° |
| R5 | Pers. PageRank | 0.014° | 0.018° | 0.061▼ | 0.082▼ |

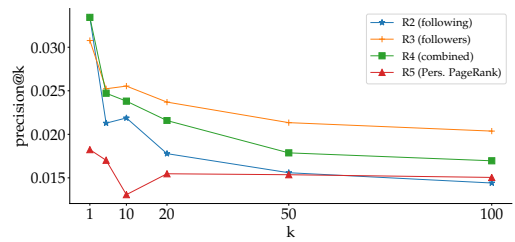

Figure 1: Precision for different recommendation list lengths ($k$) in offline evaluation.

The online evaluation is performed as follows. A recommendation bot is created on the Mastodon instance associated with the institute of the authors[2]. Afterwards, we ask users to follow this bot if they wish to receive personalized recommendations. For each participant, we generate a static web page consisting of a list of $N$ recommendations with the option to start to follow a suggested user. A link to this web page is then send to the user and we track the user interactions. A recommendation is considered relevant if the participant starts to follow a suggested user. The recommendations of two algorithms are presented using balanced-interleaving, which is a relatively inexpensive evaluation method for online experiments compared to conventional A/B testing. We refer the reader to [6] for a thorough discussion of this evaluation method.

One complication arises in the online evaluation. As an up-to-date graph is unavailable at recommendation time, such a graph has to be created. For this, we explore the vicinity of a recommendation target $u \in V$ by applying an egocentric random walk for a fixed amount of iterations. This strategy resembles the "circle-of-trust" used in the Twitter recommender system [4]. The random walk is performed as follows. At each iteration, the algorithm either transitions to a random neighbor of the current user with probability $\gamma$, or jumps back to $u$ with probability $1 - \gamma$. In our experiments, we execute the random walk for 200 iterations and set $\gamma = 0.8$. Here, we do not claim that this is the most efficient way of generating recommendations in an online setup. It is merely a way to deal with incomplete data in federated social networks.

## 3 RESULTS
### 3.1 Offline Evaluation
The collaborative filtering approach shows a consistently higher performance than a topology-based system using PageRank (see Table 2). With respect to the success at rank $k$ metric, profile-based approaches (R2–R4) have up to two times higher retrieval scores than PageRank (R5). The individual profiling strategies perform all rather similarly, which aligns with the findings in [5]. Also, a baseline system (R1) which generates recommendations by selecting 100 random users from the network topology is outperformed by a large margin. In Figure 1, it can be observed that shorter recommendation lists have a higher precision for all recommendation strategies. Precision at rank $k$ remains stable starting from a list length of $k = 50$ items. This suggests that shorter lists are to be preferred in an online scenario.

[2]See https://mastodon.utwente.nl/@Followdon

It is important to mention that the list of possible suggestions from the profile-based recommender is smaller than the list from the PageRank recommender, which complicates the discussion. Only visited nodes (see Section 2.3) have been indexed in the document retrieval system. This significantly reduces the pool size of possible users (≈3k). In contrast, the PageRank recommender can suggest any user in the topology (≈255k). One could overcome this issue as follows. Each user, regardless of whether or not it has been visited during the data collection, could be added to the search index. Then, incoming relationships can be inferred by inspecting the outgoing links of visited users. By adding these relations as *followers* to the documents of unexplored nodes, the *following* strategy of the collaborative filtering can be applied. However, as no outgoing links are known for unexplored users, the *following* and *combined* strategies are not fully applicable. Due to time constraints, we did not further investigate this issue.

Furthermore, it is worth to note that the chosen window of five days between $t_1$ and $t_2$ might not have been long enough to capture sufficient user activity. In between the training snapshot at $t_1$ and the testing snapshot at $t_2$, six new connections were added to each user on average. This gives rise to an interesting trade-off. For a longer time span, one can capture larger amounts of activity within the network. Intuitively, more links will be added as users start to follow other users. However, the farther two snapshots are apart, the larger is the risk that the network deviates too much from the original structure. Users might stop following other users or profiles could be deleted. More severely, entire instances could become unavailable due to a temporary downtime, or they could even be discontinued. This is a unique concern related to the distributed nature of federated social networks.

Finally, we want to motivate which recommendation systems are evaluated in the online trial based on the results presented above. From Table 2 it can be observed that the profile-based recommendation strategies perform rather similarly. However, the combined strategy (R4) performs best with respect to the success at rank 10 metric, which one seeks to maximize in an online system where 10 recommendations are presented to the user. Therefore, we pick R4 as the first recommendation system. Although the personalized PageRank recommender (R5) has a lower performance than the other profiling strategies, we expect that it produces valuable recommendations which are significantly different from the profile-based strategies. This is due to the fact that it considers the network topology when generating recommendations. Therefore, we apply the balanced-interleaving evaluation to systems R4 and R5.

Table 3: Summary of online evaluation.

| Characteristic | Value |
| --- | --- |
| Number of participants | 19 |
| Profile-based recommender (R4) superior | 5 |
| PageRank recommender (R5) superior | 5 |
| Draw | 2 |
| No user interaction | 7 |

### 3.2 Online Evaluation

The online evaluation shows that neither the profile-based nor the topology-based system is superior (see Table 3). Nineteen users participated in our online study. On average, they started to follow 1.8 users from our recommendations. For 5 users the profile-based approach performed best. For another 5 users, the topology-based approach performed best. For the remaining 9 users both system performed equally well, or no recommendation was followed. The fact that valuable recommendations were generated that resulted in new followings shows that the two systems can be useful in practice. However, a larger group of participants is required to draw final conclusions on the recommender system performance.

### 3.3 Practical Considerations

The generation of online recommendations turned out to be costly because the complete network data is not available. In contrast to centralized social media, federated social networks do not have a single authority which stores data about the entire network graph. The proposed method of crawling the vicinity of a target user at recommendation time (see Section 2.4) comes with a high overhead in network traffic and is not suitable for real-time systems that have to support large amounts of users. In addition to that, the method is sensitive to the size of the vicinity. We expect that a larger number of iterations yields a better picture of a user's vicinity, which in turn increases the quality of recommendations. However, an exploration of different parameter settings has been out of scope of this study. The data collection issue is even more severe in the offline evaluation which requires large and representative samples of the entire network.

To reduce the overhead associated with crawling in an online setting, one might attempt to gradually construct a cached representation of the entire network graph. Whenever a recommendation is generated for a user, the vicinity is added to that graph. On subsequent recommendations, one might reuse parts of this network to avoid additional crawling. This approach has two important issues that have to be considered. First, one has to address the question when parts of the network are considered to be out of date (i.e., when the cache expires). Second, and more importantly, such an approach seems to be in conflict with the intentions behind decentralization. By constructing a database that aims to capture the entire network graph, one starts to centralize the data of a federated social network.

## 4 CONCLUSION

User recommendation algorithms commonly applied to centralized social media can be applied to incomplete data from federated social networks with the goal of developing an engaged community. We showed that collaborative filtering-based recommenders outperform a topology-based recommender on a large unbiased sample of the federated social network Mastodon. The two recommenders outperform a random recommender by a large margin. A subsequent live user experiment on Mastodon using balanced interleaving shows that the two recommender approaches perform on par. Acquiring a sufficiently large snapshot of the network topology for offline recommendation proofed to be difficult and costly. Keeping the snapshot up-to-date needs constant re-sampling. Online recommendation was done by sampling the graph neighborhood for the current user.

There are several directions for future work. First, studying the extent to which incomplete data impacts the recommender performance may derive methods that are tailored towards federated social networks which operate with limited amounts of data. Second, user recommendation algorithms in popular social media increasingly utilize user context information such as location data and interests. It remains unclear how such data can be effectively acquired and utilized in federated social networks while preserving privacy. Third, BM25 might not be the best ranking function for the presented recommender approach, and it should be compared to functions that also use popularity-based scoring. Finally, one may investigate how decentralized communication protocols such as ActivityPub can be extended to support community building algorithms while maintaining the notion of decentralized network data.

## REFERENCES

[1] Jonathan Anderson, Claudia Diaz, Joseph Bonneau, and Frank Stajano. 2009. Privacy-enabling Social Networking over Untrusted Networks. In *Proceedings of the 2nd ACM Workshop on Online Social Networks (WOSN '09)*. 1–6.

[2] Ames Bielenberg, Lara Helm, Anthony Gentilucci, Dan Stefanescu, and Honggang Zhang. 2012. The growth of Diaspora - A decentralized online social network in the wild. *2012 Proceedings IEEE INFOCOM Workshops* (2012), 13–18.

[3] Minas Gjoka, Maciej Kurant, Carter T. Butts, and Athina Markopoulou. 2010. Walking in Facebook: A Case Study of Unbiased Sampling of OSNs. In *Proceedings of the 29th Conference on Information Communications (INFOCOM'10)*. 2498–2506.

[4] Pankaj Gupta, Ashish Goel, Jimmy Lin, Aneesh Sharma, Dong Wang, and Reza Zadeh. 2013. WTF: The Who to Follow Service at Twitter. In *Proceedings of the 22nd International Conference on World Wide Web (WWW '13)*. 505–514.

[5] John Hannon, Mike Bennett, and Barry Smyth. 2010. Recommending Twitter Users to Follow Using Content and Collaborative Filtering Approaches. In *Proceedings of the 4th ACM Conference on Recommender Systems (RecSys'10)*. 199–206.

[6] Thorsten Joachims. 2003. Evaluating retrieval performance using clickthrough data. In *Text Mining*, Jürgen Franke, Gholamreza Nakhaeizadeh, and Ingrid Renz (Eds.). Springer, 266–290.

[7] David Liben-Nowell and Jon Kleinberg. 2003. The Link Prediction Problem for Social Networks. In *Proceedings of the Twelfth International Conference on Information and Knowledge Management (CIKM '03)*. 556–559.

[8] Christopher D. Manning, Prabhakar Raghavan, and Hinrich Schütze. 2008. *Introduction to Information Retrieval*. Cambridge University Press.

[9] Lawrence Page, Sergey Brin, Rajeev Motwani, and Terry Winograd. 1999. *The PageRank Citation Ranking: Bringing Order to the Web*. Technical Report 1999-66. Stanford InfoLab.

[10] Eugen Rochko. 2018. The mastodon project. Retrieved November 14, 2018 from https://joinmastodon.org

[11] Amre Shakimov, Alexander Varshavsky, Landon P. Cox, and Ramón Cáceres. 2009. Privacy, Cost, and Availability Tradeoffs in Decentralized OSNs. In *Proceedings of the 2nd ACM Workshop on Online Social Networks (WOSN '09)*. 13–18.

[12] Tianyi Wang, Yang Chen, Zengbin Zhang, Peng Sun, Beixing Deng, and Xing Li. 2010. Unbiased Sampling in Directed Social Graph. In *Proceedings of the ACM SIGCOMM 2010 Conference (SIGCOMM '10)*. 401–402.

[13] Tianyi Wang, Yang Chen, Zengbin Zhang, Tianyin Xu, Long Jin, Pan Hui, Beixing Deng, and Xing Li. 2011. Understanding Graph Sampling Algorithms for Social Network Analysis. In *Proceedings of the 2011 31st International Conference on Distributed Computing Systems Workshops (ICDCSW '11)*. 123–128.

# From Neural Re-Ranking to Neural Ranking: Learning a Sparse Representation for Inverted Indexing

Hamed Zamani[1]  Mostafa Dehghani[2]  W. Bruce Croft[1]  Erik Learned-Miller[1]  Jaap Kamps[2]
[1]University of Massachusetts Amherst   [2]University of Amsterdam

## 1 Extended Abstract[*]

Retrieving unstructured documents in response to a natural language query is the core task in information retrieval (IR). Due to the importance of this task, the IR community has put a significant emphasis on designing efficient and effective retrieval models since the early years. The recent and successful development of deep neural networks for various tasks has also impacted IR applications. In particular, neural ranking models (NRMs) have recently shown significant improvements in a wide range of IR applications, such as ad-hoc retrieval, question answering, context-aware retrieval, mobile search, and product search. Most of the existing neural ranking models have a specific property in common: they are employed for *re-ranking* a small set of potentially relevant documents for a given query, provided by an efficient first stage ranker. In other words, since most neural ranking models rely on semantic matching that can be achieved using distributed dense representations, computing the retrieval score for all the documents in a large-scale collection is generally infeasible. Queries are short and terms have a highly skewed Zipfian distribution making each term relatively selective, resulting in a simple join over very few relatively short posting lists. In contrast, dense representations have an almost uniform distribution, with every term (to some degree) matching essentially all documents —similar to extreme stopwords that we cannot filter out.

Our approach addresses this head-on: by enforcing and rewarding sparsity in the representation learning, we create a latent representation that aims to capture meaningful semantic relations while still parsimoniously matching documents. This is illustrated in Figure 1, showing that the Zipfian distribution of the term space is matching far fewer documents than the dense representation returning collection-length posting lists for every term, dramatically increasing index size and query processing time. However, the latent sparse representation proposed in this paper mimics the posting list length distribution of the term based model, even matching fewer documents than term based models. That is, unlike existing neural ranking models, we propose to learn *high-dimensional sparse* representations for query and documents in order to allow for an inverted index based *standalone neural ranker* (SNRM). Our model does not require a first stage ranker and can retrieve documents from a large-scale collection as efficient as conventional term matching models.

Our main goal is learning representations for documents and queries that result in better matching compared to the original term vectors and exact matching models, while we still inherit the efficiency rooted in the sparsity of those representations. So there are two objectives, *introducing sparsity* and *capturing latent semantic meanings*. We first maps ngrams of queries and documents to a low-dimensional dense representation to compress the information, and then transform it to a high-dimensional representation pursuing the sparsity as a desired characteristic for these representations. By aggregating over sparse ngram representations, we obtain a sparse representation for a text with an arbitrary length, whose sparsity is a function of the input length: this implies higher sparsity for queries in comparison with documents, achieving an efficient retrieval model. We achieve a sparsity ratio in the learned representations that is comparable to the sparsity ratio in original term vectors.

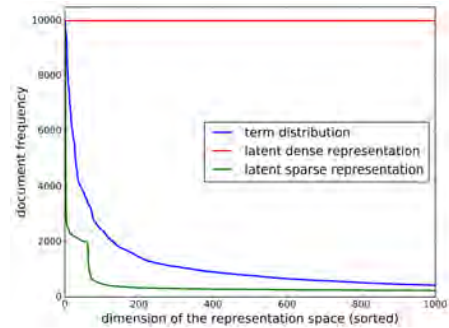

**Figure 1: Document frequency in the term space (blue), the latent dense space (red), and the latent sparse space (green).**

Once our latent sparse representation is trained offline, we initiate an inverted index construction phase that looks at each dimension of the learned representation as a "latent term" and builds an inverted index from each latent term to each document of the collection. At query time, we transform a given query to the learned latent high-dimensional space, and obtain its sparse representation. Given the small number of non-zero elements of the query representation and the constructed inverted index, we are able to retrieve documents from the entire collection efficiently. In addition, we can perform traditional pseudo-relevance feedback in the learned semantic space. We conduct extensive experiments with SNRM on TREC Robust and ClueWeb collections demonstrating the effectiveness of the proposed model. In summary, we show that SNRM gains in efficiency without loss of effectiveness: it not only outperforms the existing term matching baselines, but also performs similarly to the recent re-ranking based neural models with dense representations. Our model can also take advantage of pseudo-relevance feedback for further improvements. More generally, our results demonstrate the importance of sparsity in neural IR models and show that dense representations can be pruned effectively, giving new insights about essential semantic features and their distributions.

---

[*]This is an extended abstract of Zamani et al. [1].

## References

[1] Hamed Zamani, Mostafa Dehghani, W. Bruce Croft, Erik Learned-Miller, and Jaap Kamps. 2018. From Neural Re-Ranking to Neural Ranking: Learning a Sparse Representation for Inverted Indexing. In *CIKM'18: Proceedings of the 2018 ACM on Conference on Information and Knowledge Management*. https://doi.org/10.1145/3269206.3271800

# Aspect-based summarization of pros and cons in unstructured product reviews

Florian Kunneman
Radboud University
Nijmegen, The Netherlands
f.kunneman@let.ru.nl

Sander Wubben
Tilburg University
Tilburg, The Netherlands
s.wubben@uvt.nl

Antal van den Bosch
KNAW Meertens Institute
Amsterdam, The Netherlands
a.vandenbosch@let.ru.nl

Emiel Krahmer
Tilburg University
Tilburg, The Netherlands
e.j.krahmer@uvt.nl

## 1 SUMMARY

This work is a compression of [2].

We implemented and evaluated a rule-based and a data-driven approach to aspect-based sentiment analysis, with a focus on Dutch product reviews of electronic devices on the platform kieskeurig.nl. The rule-based system matches review texts to a set of syntactic patterns and a valency lexicon: any phrase that matches a heuristically set syntactic pattern and a positive or negative word in the Duoman lexicon [1] is extracted as a pro or con. The data-driven approach leverages the pros and cons put forward by the writers of the reviews themselves. About 30% of the users make the effort to fill in these pros and cons fields. We implemented two versions of supervised classification: a shallow neural network trained on the original pros and cons and a shallow neural network trained on clusters of the pros and cons. The second version served to reduce the variety in pros and cons that was seen (e.g.: many different expressions to communicate a similar evaluation), and was implemented as $K$-means clustering with a target of 100 clusters.

We trained, developed and tested our systems on a dataset of 4.575 reviews, more or less equally divided over four types of devices: vaccuum cleaners, deep fryers, espresso machines and smartphones. We added a baseline that matched known pros and cons from the training set to $N$-grams in the test reviews. While the pros and cons put forward by the users offers a gold standard for evaluation, we also included a human evaluation in which the qualities of both the system-generated summaries and the human-generated summaries are assessed. The results on this second evaluation are given in Table 1, where completeness is scored on a scale from 1 to 7 and relevance relates to the percentage of pros and cons that are deemed relevant to the review text. 'SynPat' is the rule-based system and the two systems with 'Neural' in the name are the supervised systems. The output of the former is clearly better than those of the latter. Even the pros and cons put forward by the humans themselves (named 'Reviewers summary') are deemed as less relevant to the review text and only slightly more complete than the ones extracted by the rule-based system. Hence, these pros and cons might not be suitable for supervised modelling of aspect-based sentiment analysis.

|  | Completeness | Relevance |
|---|---|---|
| Baseline | 3.90 (0.90) | 0.44 (0.17) |
| SynPat | 4.06 (1.15) | 0.67 (0.25) |
| Neural | 2.74 (0.86) | 0.25 (0.14) |
| Neural_clust | 2.37 (0.62) | 0.18 (0.10) |
| Reviewers summary | 4.60 (1.13) | 0.61 (0.25) |

**Table 1: Outcomes of the human assessment of the pros and cons generated by each system and the writers of the reviews themselves for 20 reviews.**

## 2 MOTIVATION

Based on the insights from this study, a system can be developed that automatically extracts pros and cons from product reviews. As only 30% of the reviews in our dataset were provided with pros and cons by the writers themselves, this system can extract the pros and cons for the remaining 70% and facilitate the writers of new reviews with suggested pros and cons. Subsequently, multiple reviews on the same product could be summarized by counting pros and cons and provide visitors of the platform with a swift overview. Finally, knowledge on the valency of consumer products can provide input to (personalized) recommender systems.

Scientifically, this study stands out in its evaluation, obtaining insight into the quality of gold standard labels by evaluating the pros and cons suggested by several systems alongside the pros and cons that the writers themselves have filled in. This could form the ground for a discussion of the quality of distantly supervised labels.

## REFERENCES

[1] Valentin Jijkoun and Katja Hofmann. 2009. Generating a non-english subjectivity lexicon: Relations that matter. In *Proceedings of the 12th Conference of the European Chapter of the Association for Computational Linguistics.* Association for Computational Linguistics, 398–405.
[2] Florian Kunneman, Sander Wubben, Antal van den Bosch, and Emiel Krahmer. 2018. Aspect-based summarization of pros and cons in unstructured product reviews. In *Proceedings of the 27th International Conference on Computational Linguistics.* 2219–2229.

# Search as a learning activity: a viable alternative to instructor-designed learning?

Felipe Moraes
Delft University of Technology
Delft, The Netherlands
f.moraes@tudelft.nl

Sindunuraga Rikarno Putra
Delft University of Technology
Delft, The Netherlands
sindunuragarikarnoputra@student.tudelft.nl

Claudia Hauff
Delft University of Technology
Delft, The Netherlands
c.hauff@tudelft.nl

Search and sensemaking is an intricate part of the learning process, and for many learners today synonymous with accessing and ingesting information through Web search engines [1, 6, 9]. At the same time, Web search engines are not built to support users in the type of complex searches often required in learning situations [2–4]. But what effect does this lack of a learning-focused Web search engine design have on the ability of users to learn compared to a setting where they are provided with high-quality learning materials? In this paper we set out to answer this question by *measuring* how effective searching to learn is compared to (i) learning from—in our experiment: high-quality video—materials specifically designed for the purpose of learning, (ii) learning from video materials in combination with search, and, (iii) searching together with a partner to learn (i.e. collaborative search for learning).

The aim of our work is to *quantify* to what extent search as a learning activity is a viable alternative to what we call *instructor-designed learning*, that is, learning materials designed and created specifically for the purpose of learning. As not for every possible topic specifically designed learning materials exist, it is important to understand what effect that has on one's ability to learn. In addition, we are also interested in understanding whether the lack of learning materials can be compensated in the search setting by the presence of a second learner that has the same learning intent (i.e. collaborative search for learning).

Our work is guided by the following research questions:

**RQ1** How effective (with respect to learning outcome) is searching to learn compared to instructor-designed learning?
**RQ2** How effective (with respect to learning outcome) is instructor-designed learning *supported by search* in comparison to just instructor-designed learning?
**RQ3** How effective is pair-wise collaborative search compared to single-user search for learning?

Specifically, in this work we conducted a user study with 151 participants and measured *vocabulary learning*, a particular instance of human learning (similar in spirit to [7, 8]), across five search and instructor-designed learning conditions (Figure 1 depicts our search conditions). As high-quality instructor-designed learning materials we make use of lecture videos sourced from TED-Ed, Khan Academy and edX, popular online learning platforms.

---

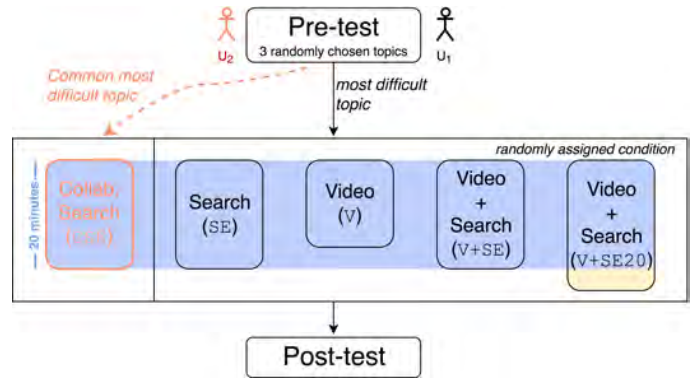

**Figure 1: Study design overview: four single-user conditions and one collaborative (pairwise) condition.**

## FINDINGS

Our main findings can be summarised as follows:

- we find participants in the instructor-designed learning condition (watching high-quality lecture videos) to have 24% higher learning gains than participants in the searching to learn condition;
- collaborative search as learning does not result in increased learning gains;
- the *combination* of instructor-designed learning and searching to learn leads to significantly higher learning gains (an increase of up to 41%) than the instructor-designed learning condition without a subsequent search phase.

## REFERENCES
[1] J Patrick Biddix, Chung Joo Chung, and Han Woo Park. 2011. Convenience or credibility? A study of college student online research behaviors. *The Internet and Higher Education* (2011), 175–182.
[2] Gene Golovchinsky, Abdigani Diriye, and Tony Dunnigan. The future is in the past: designing for exploratory search. In *IIiX'2012*. 52–61.
[3] Ahmed Hassan Awadallah, Ryen W White, Patrick Pantel, Susan T Dumais, and Yi-Min Wang. 2014. Supporting complex search tasks. In *CIKM '14*. 829–838.
[4] Gary Marchionini. 2006. Exploratory search: from finding to understanding. *Comm. ACM* (2006), 41–46.
[5] Felipe Moraes, Sindunuraga Rikarno Putra, and Claudia Hauff. 2018. Contrasting Search as a Learning Activity with Instructor-designed Learning. (2018).
[6] David Nicholas, Ian Rowlands, David Clark, and Peter Williams. 2011. Google Generation II: web behaviour experiments with the BBC. In *ASLIB*. 28–45.
[7] Rohail Syed and Kevyn Collins-Thompson. 2017. Optimizing search results for human learning goals. *IRJ* (2017), 506–523.
[8] Rohail Syed and Kevyn Collins-Thompson. 2017. Retrieval algorithms optimized for human learning. In *SIGIR '17*. 555–564.
[9] Arthur Taylor. 2012. A study of the information search behaviour of the millennial generation. *Information Research: An International Electronic Journal* (2012).

# SearchX: Collaborative Search System for Large-Scale Research

Sindunuraga Rikarno Putra, Kilian Grashoff
Delft University of Technology
Delft, The Netherlands
{sindunuragarikarnoputra,k.c.grashoff}@student.tudelft.nl

Felipe Moraes, Claudia Hauff
Delft University of Technology
Delft, The Netherlands
{f.moraes,c.hauff}@tudelft.nl

Web search is generally seen as a solitary activity, as most mainstream technologies are designed for single-user search sessions. However, for a sufficiently complex task, collaboration during the information seeking process is beneficial [4]. A survey by Morris [6] has shown that collaborating during search is a common activity, albeit using ad hoc solutions such as email and instant messaging. Morris also found a significant increase in the number of people who collaborate during search at a regular basis, from 0.9% in 2006 to 11% in 2012. This increasing use of collaborative search (CSE) has also been reflected in the research community, where CSE has been an active area of research for many years. Workshops that explicitly focus on collaborative search—and more generally information seeking—have started to appear in 2008 [9] and continue to do so to this day, as for example by Azzopardi et al. [2].

In contrast to single-user search where a number of up-to-date and open-source tools are readily available (e.g. Terrier[1] and Elasticsearch[2]), the CSE research community has currently just one maintained open-source option (Coagmento) despite the fact that researchers have designed and implemented a number of systems in the past ten years [1, 3, 5, 7, 8]. While Coagmento provides an extensive collaboration feature set, it requires users to either install a browser plugin or an Android/iOS app, making it less viable for large-scale CSE experiments which are often conducted with crowd workers. Furthermore, we believe as researchers we should have a choice of tooling, instead of relying on a single one.

For these reasons, we have designed and implemented SearchX, a CSE system built on modern Web standards, allowing it to be accessed from multiple platforms without the need for user-side installations. We designed SearchX specifically for CSE research and provide a comprehensive documentation to enable others to implement and run their own CSE experiments.

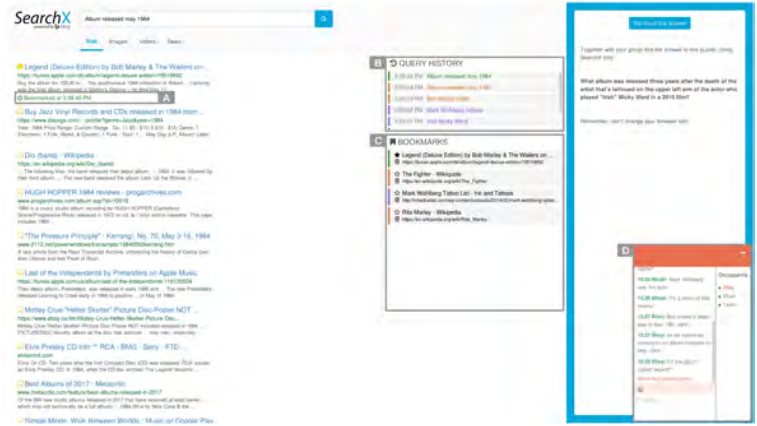

**Figure 1: SearchX Collaborative search interface. [A] bookmarking information including who bookmarked, [B] shared query history, [C] shared bookmarks, [D] chat.**

## REFERENCES
[1] Saleema Amershi and Meredith Ringel Morris. 2008. CoSearch: A System for Co-located Collaborative Web Search. In *CHI '08*. 1647–1656.
[2] Leif Azzopardi, Jeremy Pickens, Chirag Shah, Laure Soulier, and Lynda Tamine. 2017. Second International Workshop On the Evaluation of Collaborative Information Seeking and Retrieval (Ecol'17). In *CHIIR '17*. 429–431.
[3] Robert Capra, Annie T. Chen, Katie Hawthorne, Jaime Arguello, Lee Shaw, and Gary Marchionini. 2012. Design and evaluation of a system to support collaborative search. *ASIST* 49, 1 (2012), 1–10.
[4] Jonathan Foster. 2006. Collaborative Information Seeking and Retrieval. *Annual Review of Information Science and Technology* 40, 1 (Dec. 2006), 329–356.
[5] Gene Golovchinsky, John Adcock, Jeremy Pickens, Pernilla Qvarfordt, and Maribeth Back. 2008. Cerchiamo: a collaborative exploratory search tool. *Computer Supported Cooperative Work* (2008), 8–12.
[6] Meredith Ringel Morris. 2013. Collaborative Search Revisited. In *CSCW '13*. 1181–1192.
[7] Meredith Ringel Morris and Eric Horvitz. 2007. SearchTogether: An Interface for Collaborative Web Search. In *UIST '07*. 3–12.
[8] Sharoda A. Paul and Meredith Ringel Morris. 2009. CoSense: Enhancing Sensemaking for Collaborative Web Search. In *CHI '09*. 1771–1780.
[9] Jeremy Pickens, Gene Golovchinsky, and Meredith Ringel Morris. 2009. Proceedings of 1st International Workshop on Collaborative Information Seeking. *CoRR* abs/0908.0583 (2009).
[10] Grashof Kilian Putra, Sindunuraga Rikarno, Felipe Moraes, and Claudia Hauff. On the Development of a Collaborative Search System. In *DESIRES'18*.
[11] Sindunuraga Rikarno Putra, Felipe Moraes, and Claudia Hauff. SearchX: Empowering Collaborative Search Research.. In *SIGIR'18*.

---

*This is a compressed version of Putra et al. [10, 11].
[1] http://terrier.org/
[2] https://www.elastic.co/

---

# Narrative-Driven Recommendation as Complex Task

Toine Bogers
Aalborg University Copenhagen
Denmark
toine@hum.aau.dk

Marijn Koolen
Royal Netherlands Academy of Arts and Sciences
Netherlands
marijn.koolen@di.huc.knaw.nl

## ABSTRACT
This paper is an extended abstract of [2] and [3].

## 1 INTRODUCTION
Current-generation recommendation algorithms are often focused on generic ratings prediction and item ranking tasks based on a user's past preferences. However, many scenarios are more complex with specific criteria and constraints on which items are relevant. This paper focuses on a particular type of complex recommendation needs: Narrative-Driven Recommendation (NDR), where users describe their needs in short narratives, often with one or more example items that fit that need, against a background of historical preferences that may not be spelled out in the narrative, but do play a role in their considerations. We show that such complex needs are common on the Web, yet current-generation systems offer limited to no support for these needs. We focus on narrative-driven book recommendation in the context of LibraryThing (LT) users posting recommendation requests in the discussion forums. We provide an analysis of these needs in terms of their structure, the relevance aspects they cover, and what types of data and algorithms fits these aspects. Subsequently, we propose several new algorithms that take advantage of these narratives and example items as well as hybrid systems, most of which significantly outperform classic collaborative filtering. We show that NDR is indeed a complex scenario that requires further study. Our findings have consequences for system design and development not only in the book domain, but also in other domains where users express focused recommendation needs, such as movies, television, games and music.

## 2 REQUEST ANALYSIS
We analysed a random sample of discussion threads on the LT forums and found that 9% of these focus on recommendation requests, making this a prevalent recommendation need and scenario. Of these requests, 58% contain example books or authors. The narrative often contains content-related criteria such as topic, genre, style and difficulty level, but also familiarity aspects based on previous reading experiences (e.g. books with similar style, mood or plot as some given example books). Other important relevance aspects are engagement, accessibility and metadata (e.g. books by a certain author or by a specific publisher). Similar types of NDR requests were found in other domains such as games, movies and music [1]. The different relevance aspects require different types of data and different types of relevance models. E.g. familiarity aspects require a user's personal preference information (transactions and ratings) and latent factor analysis. Topical content aspects require subject analysis as found in library metadata and user tags and reviews.

## 3 EVALUATION
The user's recommendation need is partially represented by the narrative request and the example books and authors. Using the narrative request as a textual representation is a form of **Narrative-Driven Recommendation** (NDR). Using the example books and authors as a mini-profile is referred to as Example-Driven Recommendation (EDR).

We extended the test collection from the CLEF 2016 Social Book Search Lab [4] with additional requests and relevance judgments, and evaluate a number of standard content-based filtering (CBF) and collaborative filtering (CF) approaches, using both NDR and EDR. We find that NDR is more effective than EDR and than traditional matrix factorization using the entire user profile, especially when using user-generated content such as reviews and tags for matching. Example books are more effective than example authors, probably due to the author representation leading to topic drift. However, combining NDR and EDR in a hybrid system significantly outperforms each individual approach, showing the narrative and examples provide complementary signals.

## 4 CONCLUSIONS
Narrative-Driven recommendation is a complex scenario that requires multiple data sources and algorithms to solve, and exploiting user-generated content is essential to good performance. Future work includes semantic analysis of the requests to extract structured information, exploring the value of conversational search and recommendation models and knowledge-aware approaches and testing in the domains of games, movies and music.

## REFERENCES
[1] Toine Bogers, Maria Gäde, Marijn Koolen, Vivien Petras, and Mette Skov. "What was this Movie About this Chick?" A Comparative Study of Relevance Aspects in Book and Movie Discovery. In *Proc iConf 2018*. Vol. 10766. Springer, 323–334.
[2] Toine Bogers and Marijn Koolen. 2017. Defining and Supporting Narrative-driven Recommendation. In *RecSys '17: Proceedings of the Eleventh ACM Conference on Recommender Systems*. ACM, 238–242.
[3] Toine Bogers and Marijn Koolen. 2018. "I'm looking for something like ...": Combining Narratives and Example Items for Narrative-driven Book Recommendation. In *KARS '18: Proceedings of the First Knowledge-aware and Conversational Recommender Systems Workshop*. CEUR-WS.
[4] Marijn Koolen, Toine Bogers, Maria Gäde, Mark M. Hall, Iris Hendrickx, Hugo C. Huurdeman, Jaap Kamps, Mette Skov, Suzan Verberne, and David Walsh. 2016. Overview of the CLEF 2016 Social Book Search Lab. In *Experimental IR Meets Multilinguality, Multimodality, and Interaction - 7th International Conference of the CLEF Association, CLEF 2016, Évora, Portugal, September 5-8, 2016, Proceedings*. 351–370.

# Measuring User Satisfaction on Smart Speaker Intelligent Assistants

Seyyed Hadi Hashemi*
University of Amsterdam
Amsterdam, The Netherlands
hashemi@uva.nl

Kyle Williams
Microsoft
Redmond, USA
Kyle.Williams@microsoft.com

Ahmed El Kholy
Microsoft
Redmond, USA
Ahmed.ElKholy@microsoft.com

Imed Zitouni
Microsoft
Redmond, USA
izitouni@microsoft.com

Paul A. Crook[†]
Facebook
Seattle, USA
pacrook@fb.com

## 1 INTRODUCTION

Intelligent assistants are increasingly being used on smart speaker devices, such as Amazon Echo, Google Home, Apple Homepod, and Harmon Kardon Invoke with Cortana. Typically, user satisfaction measurement relies on user interaction signals, such as clicks and scroll movements, in order to determine if a user was satisfied. However, these signals do not exist for smart speakers, which creates a challenge for user satisfaction evaluation on these devices. In order to address measuring user satisfaction on smart speaker Intelligent Assistant (IA), we have studied task and session identification on smart speaker IAs [1] and user satisfaction measurement on smart speaker IAs using intent sensitive query embeddings [2].

## 2 TASK AND SESSION IDENTIFICATION

Task and session identification is a key element of system evaluation and user behavior modeling in IA systems. However, identifying task and sessions for IAs is challenging due to the multi-task nature of IAs and the differences in the ways they are used on different platforms, such as smart-phones, cars, and smart speakers. Considering the multi-task nature of the users' behaviors in IA, we follow the task and session definitions as proposed in [1]:

- **A Task** is a single information need that can be satisfied by at least one query and one IA generated response.
- **A Session** is a short period of contiguous time spent to fulfill one or multiple tasks.

Furthermore, usage behavior may differ among users depending on their expertise with the system and the tasks they are interested in performing. In this study, we investigate how to identify tasks and sessions in IAs given these differences. To this aim, we analyze data based on the interaction logs of two IAs integrated with smart-speakers. We fit Gaussian Mixture Models to estimate task and session boundaries and show how a model with 3 components models user interactivity time better than a model with 2 components. We then show how session boundaries differ for users depending on whether they are in a learning-phase or not. Finally, we study how user inter-activity times differs depending on the task that the user is trying to perform. Our findings show that there is no single task or session boundary that can be used for IA evaluation. Instead, these boundaries are influenced by the experience of the user and the task they are trying to perform.

## 3 USER SATISFACTION MEASUREMENT

Given the definitions of tasks and session, we define task satisfaction as follows:

- **Task satisfaction** is how successful a user is in completing a single information need using at least one query and receiving at least one IA generated response.

In this study, we propose a new signal, user intent, as a means to measure user satisfaction. We propose to use this signal to model user satisfaction in two ways: 1) by developing intent sensitive word embeddings and then using sequences of these intent sensitive query representations to measure user satisfaction; 2) by representing a user's interactions with a smart speaker as a sequence of user intents and thus using this sequence to identify user satisfaction. Our experimental results indicate that our proposed user satisfaction models based on the intent-sensitive query representations have statistically significant improvements over several baselines in terms of common classification evaluation metrics. In particular, our proposed task satisfaction prediction model based on intent-sensitive word embeddings has a 11.81% improvement over a generative model baseline and 6.63% improvement over a user satisfaction prediction model based on Skip-gram word embeddings in terms of the F1 metric. Our findings have implications for the evaluation of Intelligent Assistant systems.

## REFERENCES

[1] Seyyed Hadi Hashemi, Kyle Williams, Ahmed El Kholy, Imed Zitouni, and Paul Crook. 2018. Impact of Domain and User's Learning Phase on Task and Session Identification in Smart Speaker Intelligent Assistants. In *Proceedings of the 27th ACM International Conference on Information & Knowledge Management.*
[2] Seyyed Hadi Hashemi, Kyle Williams, Ahmed El Kholy, Imed Zitouni, and Paul Crook. 2018. Measuring User Satisfaction on Smart Speaker Intelligent Assistants Using Intent Sensitive Query Embeddings. In *Proceedings of the 27th ACM International Conference on Information & Knowledge Management.*

---

*Work done while interning at Microsoft.
[†]Work done while at Microsoft.

# WASP: Web Archiving and Search Personalized
## Compressed Contribution

Arjen P. de Vries
Radboud University
Nijmegen, The Netherlands
arjen@cs.ru.nl

## ABSTRACT
This compressed contribution to DIR 2018 introduces WASP (Web Archiving and Search Personalized), a fully functional prototype of a personal web archive and search system, which is available open source and as an executable Docker image. The original WASP paper [1] has been published at the (new, biennial) Design of Experimental Search & Information REtrieval Systems conference (DESIRES 2018).

## CCS CONCEPTS

• **Information systems** → **Information retrieval**; *Personalization*; *Search engine architectures and scalability*;

## KEYWORDS
Personal Web Archives, Personal Information Management

**ACM Reference Format:**
Arjen P. de Vries. 2018. WASP: Web Archiving and Search Personalized: Compressed Contribution. In *Proceedings of Dutch-Belgian Information Retrieval Workshop (DIR2018)*. ACM, New York, NY, USA, 1 page.

## COMPRESSED CONTRIBUTION
DESIRES is the new, systems-oriented biennial conference, focused on innovative technological aspects of search and retrieval systems, pitched as the 'CIDR for IR'. This compressed contribution shares our work on personal Web archiving and search (as published at DESIRES [1]) with the DIR community.

## MOTIVATION
Search history logs contain highly privacy-sensitive information about our interests and behaviours, online and offline, and even our health; however, 'we' have become used to handing over that very personal information to large multinational corporations, for free – in exchange for the search engine result page. With the continuing growth of compute power within reach of 'normal' individuals, we should however ask ourselves whether we really need to carry out all our searches in the cloud, on externally managed search indexes. Cannot we create our own search index and use that instead, solving the privacy concerns associated with search through a small investment in the hardware necessary to run our own search engine?

## WASP
WASP (Web Archiving and Search Personalized) provides an initial step toward claiming back our personal information, enabling us to take the responsibility for satisfying our re-finding needs into our own hands. The idea of WASP is grounded in the observation that the Web browser serves as the primary interface to access our digital information. WASP provides the tools to capture one's personal Web browsing history into a *personal* Web archive and offers a powerful retrieval interface over that history. This browser-focused setup enables the user to recall information they personally gathered without the need to deal with the large variety of information sources.

In addition to a detailed technical description of the WASP prototype, the paper reports on the observations that we made upon using the software ourselves, including an error analysis regarding archiving quality. The paper also includes a discussion of the challenges for personal web archiving and search identified while using the WASP prototype — provided both open source and as an executable Docker container so that others can use it within their research or personal life-logging setup.[1,2]

The (poster) presentation at DIR presents the key contributions of the DESIRES paper, including a preview of extensions that are under development to let WASP evolve into my ideal of the *Personal Search Engine*.

## REFERENCES

[1] Johannes Kiesel, Arjen P. de Vries, Matthias Hagen, Benno Stein, and Martin Potthast. 2018. WASP: Web Archiving and Search Personalized. In *Proceedings of the First Biennial Conference on Design of Experimental Search & Information Retrieval Systems, Bertinoro, Italy, August 28-31, 2018. (CEUR Workshop Proceedings)*, Omar Alonso and Gianmaria Silvello (Eds.), Vol. 2167. CEUR-WS.org, 16–21. http://ceur-ws.org/Vol-2167/paper6.pdf

---

[1] https://hub.docker.com/r/webis/wasp/
[2] https://github.com/webis-de/wasp

# Melodic Similarity and Applications Using Biologically-Inspired Techniques

Dimitrios Bountouridis*, Daniel G. Brown†, Frans Wiering*, Remco C. Veltkamp†
*Department of Information and Computing Sciences, Utrecht University
†David R. Cheriton School of Computer Science, University of Waterloo
d.bountouridis@uu.nl,dan.brown@uwaterloo.ca,{f.wiering,r.c.veltkamp}@uu.nl

## ABSTRACT

Melodic similarity is a complex concept that manifests itself in a number of Music Information Retrieval (MIR) tasks such as query-by-humming and cover song detection. Typically, similarity models are based on intuition or heuristics; thus, applicability to broader contexts cannot be guaranteed. We argue that data-driven tools and analysis methods, applied to melodies known to be related, can provide us with information regarding the fine-grained nature of music similarity. Interestingly, melodic and biological sequences share a number of parallel concepts; from the natural sequence-representation, to their mechanisms of generating variations, i.e., oral transmission and evolution respectively. As such, there is a great potential for applying data-driven scientific methods and tools from bioinformatics to music. Our paper relies on such methods to *a)* acquire new knowledge through a melodic stability analysis and *b)* model global melodic similarity and apply it to a retrieval/classification scenario.

**ACM Reference Format:**
Dimitrios Bountouridis[1], Daniel G. Brown[2], Frans Wiering[1], Remco C. Veltkamp[2]. 2018. Melodic Similarity and Applications Using Biologically-Inspired Techniques: . In *Proceedings of Dutch-Belgian Information Retrieval Workshop (DIR2018)*. ACM, New York, NY, USA, 1 page.

## 1 SUMMARY

The current digital age allows listeners to stream massive collections of music. In addition, the proliferation of music streaming services has raised the listeners' interest in the accompaniment chords, the lyrics, the original versions of a cover, and many more scenarios that service providers cannot deal with manually. This development brings Music Information Retrieval (MIR) to the centre of attention. The field includes research about accurate and efficient computational methods, applied to various music retrieval and classification tasks.

Such tasks require us to build representations of previously seen classes (e.g., sets of covers of the same song), which can be only compared to a query (e.g., a cover song whose original is unknown) by means of a meaningful music similarity function. A robust MIR system should model the fuzziness and uncertainty of the differences between two musical items perceived as similar. However, this "knowledge", the exact mechanics of perceived similarity, is still unknown or incomplete. This is not surprising considering music's inherently complex nature.

To overcome, or avoid addressing the aforementioned issues, many MIR approaches to similarity rely on cognition studies, expert heuristics, music theory or formalized models in general. However, all such approaches have limited explanatory power and fail to generalize. In contrast, this paper focuses on acquiring knowledge on music and melodic similarity in particular from the data itself. Since data-driven methods and tools have been under development for years in bioinformatics, and since biological and music sequence share resembling concepts, we investigate their applicability inside a musical context.

First, this paper tackles the concept of meaningful and musically significant alignments of related melodies, by applying the bioinformatics structural alignment metaphor to music motifs. Our results reveal that the Mafft multiple alignment algorithm, which uses gap-free sections as anchor points, is a natural fit for multiple melodic sequences; a strong indication of the importance of musical patterns for melodic similarity. Trusted alignments using Mafft allow to organize melodic variations such that melodic stability/variation can be analysed. We therefore present a stability analysis free of heuristics or biases that might have been introduced following other approaches.

Secondly, this paper investigates the modelling of global melodic similarity. We capture the probability of one note to be changed to another in a variation and create musically appropriate note-substitution scoring matrices for melodic alignment. We then put these matrices successfully to the test by designing retrieval and classification tasks. Our data-driven modelling of music similarity outperforms the naive ±1 matrix, indicating that indeed some novel knowledge is captured. Additionally, we show that variations inside a melody can be an alternative source for modelling the similarity of variations among tune families or cliques of covers.

In general, we show that bioinformatics tools and methods can find successful application in music, to answer in a reliable, data-driven way a number of important, on-going questions in MIR. We argue that data-driven approaches, such as ours, constitute an ideal balance between the two occasionally contradicting goals of MIR, problem solving and knowledge acquisition.

# The Patient Forum Miner: Text Mining for patient communities

Maaike de Boer
TNO
the Hague, The Netherlands
maaike.deboer@tno.nl

Anne Dirkson
Leiden Institute for Advanced Computer Science
Leiden University
Leiden, The Netherlands
a.r.dirkson@liacs.leidenuniv.nl

Gerard van Oortmerssen
Leiden Institute for Advanced Computer Science
Leiden University
Leiden, The Netherlands
gerard.vanoortmerssen@gmail.com

Suzan Verberne
Leiden Institute for Advanced Computer Science
Leiden University
Leiden, The Netherlands
s.verberne@liacs.leidenuniv.nl

## ABSTRACT

In this demo paper we present the Patient Forum Miner (PFM), a tool for searching and browsing the archives of patient discussion groups. The PFM combines free text search with medical entity recognition and thread summarization. The demo version contains an index of 115,861 threads from the public Viva forum. We are currently indexing the forum archives of four patient communities.

**ACM Reference Format:**
Maaike de Boer, Anne Dirkson, Gerard van Oortmerssen, and Suzan Verberne. 2018. The Patient Forum Miner: Text Mining for patient communities. In *Proceedings of Dutch-Belgian Information Retrieval Workshop (DIR2018)*. ACM, New York, NY, USA, 2 pages.

## 1 INTRODUCTION

User-generated content in online forum communities is a valuable source of information: Community members can profit from the information shared in the forum, if they can retrieve the previously posted information. The experiential knowledge shared in patient communities can also be of relevance as input for medical research: There are cases of patient communities that have transformed their disease experiences into novel research data in the hope to drive research and improve their quality of life [3].

In order to make the information shared in online patient forums available to patients and medical researchers, the data needs to be indexed and made accessible through a web interface. In addition, for the purpose of knowledge discovery by experts, more structured analysis of the textual data in the forum is necessary. For example, one particular use of shared patient experiences is the detecting of adverse drug reactions (ADRs) of prescription drugs. Such post-approval monitoring is necessary due to the inability of clinical trials to fully assess the consequences of a drug before it is released to the market [2]. Open knowledge discovery such as the detection of ADRs requires the identification of relevant entities in text (diseases, medications, body parts, side effects). This task is called medical entity recognition [1].

We face the following challenges in disclosing the information contained in patient forums through free text search and medical entity recognition:

(1) **Sparse data:** The available forum data for a patient community is relatively small (compared to the web, where there is an abundance of information), which causes the findability of specific items to be low (the so-called vocabulary gap);
(2) **Noisy data:** The user-generated content on the forum is sometimes noisy, with frequent spelling errors and naming variants;
(3) **Long threads:** Threads on a single topic can easily comprise dozens or hundreds of individual posts, which makes it difficult to find the relevant information in the thread.

This has motivated us to develop text mining modules for disclosing the information in forum communities, combining free text search with query suggestion, medical entity recognition, and thread summarization. The resulting **Patient Forum Miner (PFM)** has a user interface that stimulates interactive searching and browsing behaviour [4].

The demo application at DIR[1] includes data from the large, openly available Viva forum (http://forum.viva.nl) to show the functionality of the PFM. We indexed a subset of 115,861 threads about cancer.

## 2 FUNCTIONALITY OF THE PFM

The graphical user interface of the system allows for an iterative search process in which the user quickly reaches relevant search results, supported by query suggestion, medical entity recognition, and automatic thread summarization. Figure 1 shows the system's GUI. It is divided in two main parts, the left part supporting the querying process, the right part for browsing search results. The user typically starts with entering one or more keywords (upper left). In order to accommodate for vocabulary gap problems caused by data sparsity (challenge 1), the system presents potentially relevant expansion terms to the user, generated using a Word2Vec model trained on all cancer patient forum data. By clicking one of the terms, it is added to the query.

---
[1]Restricted access through https://hematon.tnodatalab.nl

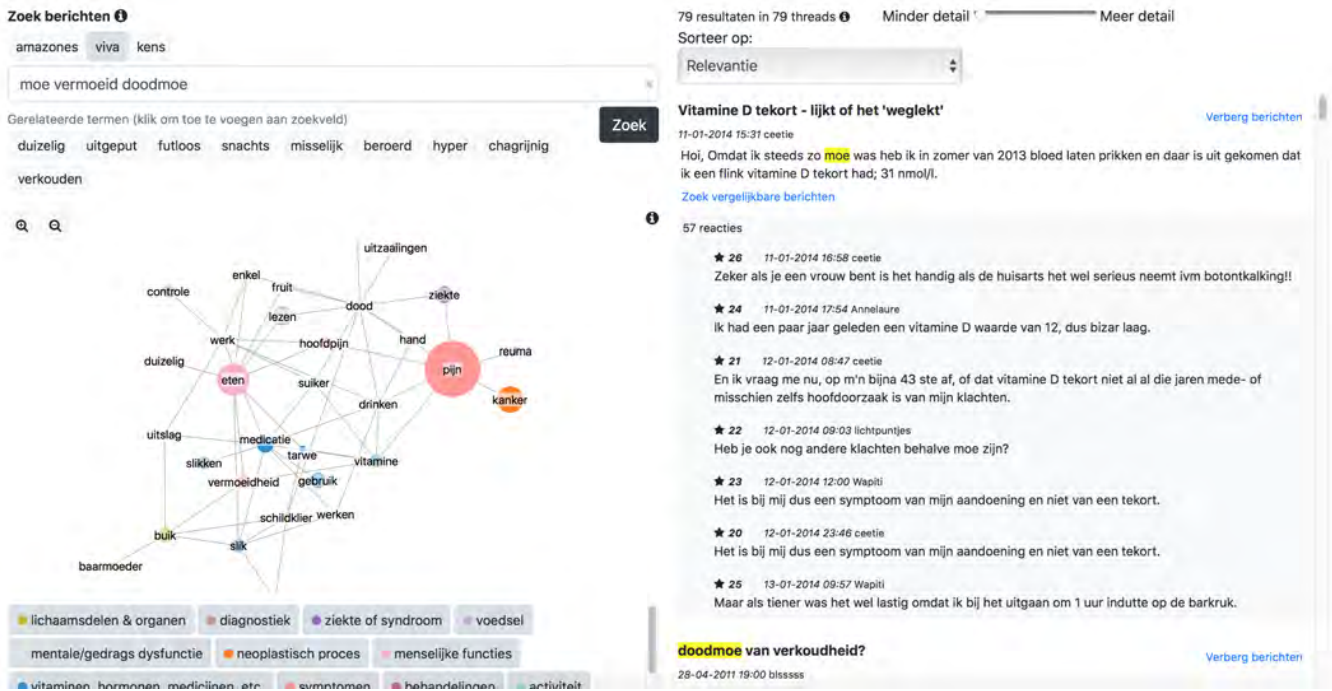

Figure 1: The user interface of the Patient Forum Miner.

The network below the query field shows the medical entities that occur in the search results and their inter-connections. The medical entities have been identified in the text through word lookup in selected categories from the Unified Medical Language System (UMLS), DBPedia, and the Medical Subject Headings (MeSH) database. For medical entity recognition from user-generated content we use the pre-processing pipelines developed in student projects [6, 7]. We specifically address issues with spelling errors by finding higher-frequent alternatives for low-frequent words with a small Levenshtein distance (challenge 2). In the presented network, two terms are connected when they co-occur frequently in the same context (i.e. message) in the result set. High-frequency terms (such as 'sleep' or 'live') were excluded from the entities, since tagging these terms was not relevant for the application.

The right hand side of the GUI shows the retrieved posts in the context of threads. We address challenge 3 through automatic summarization of the threads, showing only the most important posts (based on length, position and information quality), and from those posts only the most important sentences. The opening post of the thread is always shown, as well as posts that match the user query. If the user prefers to see more detail, they can show more sentences and posts by moving the slider on top of the screen. For details about the summarization module see Verberne et al. [5].

## 3 CURRENT AND FUTURE DEVELOPMENTS

In a project together with representatives of Dutch cancer patient communities, we evaluated the usability of the Patient Forum Miner during three meetings in the course of 2018. We are currently in the process of indexing the communities' forum archives and storing the index locally at their own web servers. In addition, we collected a number of recommendations for future developments of the PFM. One wish of the patient communities is to accommodate open knowledge discovery from their forum archives. This is a challenge that is currently addressed in a PhD project.

## ACKNOWLEDGMENTS

The authors wish to thank the volunteers of the collaborating patient organizations for their contributions to the development of the Patient Forum Miner.

## REFERENCES

[1] Asma Ben Abacha and Pierre Zweigenbaum. 2011. Medical entity recognition: A comparison of semantic and statistical methods. In *Proceedings of BioNLP 2011 Workshop*. Association for Computational Linguistics, 56–64.
[2] R Harpaz, W DuMouchel, N H Shah, D Madigan, P Ryan, and C Friedman. 2012. Novel Data-Mining Methodologies for Adverse Drug Event Discovery and Analysis. *Clinical Pharmacology & Therapeutics* 91, 6 (6 2012), 1010–1021. https://doi.org/10.1038/clpt.2012.50
[3] Ginger R. Polich. 2012. Rare disease patient groups as clinical researchers. *Drug Discovery Today* 17, 3-4 (2012), 167–172. https://doi.org/10.1016/j.drudis.2011.09.020
[4] Gerard van Oortmerssen, Stephan Raaijmakers, Maya Sappelli, Erik Boertjes, Suzan Verberne, Nicole Walasek, and Wessel Kraaij. 2017. Analyzing cancer forum discussions with text mining. *Knowledge Representation for Health Care Process-Oriented Information Systems in Health Care Extraction & Processing of Rich Semantics from Medical Texts* (2017), 127.
[5] Suzan Verberne, Antal van den Bosch, Sander Wubben, and Emiel Krahmer. 2017. Automatic summarization of domain-specific forum threads: collecting reference data. In *Proceedings of The ACM SIGIR Conference on Human Information Interaction & Retrieval (CHIIR)*. 253–256.
[6] Nicole Walasek. 2016. *Medical Entity Extraction on Dutch forum data in the absence of labeled training data*. Technical Report. TNO, the Hague, and Radboud University, Nijmegen, the Netherlands.
[7] Yiyu Yuan. 2017. *Medical Entity Recognition from Patient Forum Data*. Master's thesis. Leiden University, Leiden, the Netherlands.

# SMART Radio: Personalized News Radio

Maya Sappelli
FD Mediagroep
Amsterdam, The Netherlands
maya.sappelli@fdmediagroep.nl

Dung Manh Chu
FD Mediagroep
Amsterdam, The Netherlands
dung.manh.chu@fdmediagroep.nl

Bahadir Cambel
FD Mediagroep
Amsterdam, The Netherlands
bahadir.cambel@fdmediagroep.nl

Joeri Nortier
FD Mediagroep
Amsterdam, The Netherlands
joeri.nortier@fdmediagroep.nl

David Graus
FD Mediagroep
Amsterdam, The Netherlands
david.graus@fdmediagroep.nl

## ABSTRACT

In this demonstration paper we describe the SMART Radio app for BNR Nieuwsradio.[1] The SMART Radio app is an extension to the current BNR app, which offers users a more personalized news radio experience. It does so by automatically fragmenting shows to offer our users more targeted and focused fragments of audio, not full shows. We employ audio segmentation and audio topic-tagging techniques to achieve this, which we describe in this paper. In its present form, users can subscribe to tags to get appropriate suggestions of relevant radio fragments. In the future we would like to improve the app's personalization, by using information of the user's interaction with the app.

## CCS CONCEPTS

• **Information systems** → **Personalization**; **Recommender systems**;

## KEYWORDS

personalization, news, radio

## 1 INTRODUCTION

FD Mediagroep (FDMG[2]) is the leading information provider in the financial economic domain in the Netherlands. FDMG operates "Het Financieele Dagblad," (FD) a daily financial newspaper, similar to the Financial Times. In addition, FDMG operates the daily all-news radio station "Business News Radio" (BNR). FDMG is investing in personalizing the news experience, both in news articles and news radio [3], since we believe this will help to serve our users better, and help in unlocking long tail content from FDMG's rich archives.

[1] http://www.bnr.nl
[2] http://fdmediagroup.com

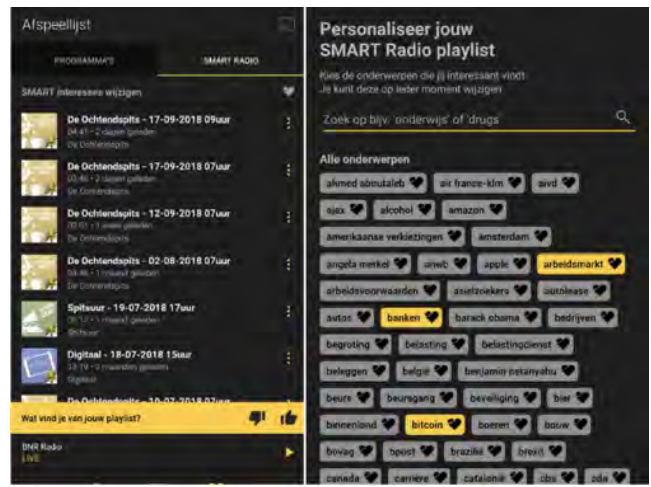

**Figure 1: Screenshot of SMART Radio in the BNR app**

Recently FDMG has launched a new version of the SMART Radio app[3] where users can create a personalized playlist, based on their interests, see also Figure 1. BNR SMART Radio offers a non-linear radio experience with short radio fragments that match the listener's interests. For this purpose, FDMG employs automatic audio segmentation, and topic-tagging techniques, which we describe in this paper.

## 2 AUDIO SEGMENTATION

On a daily basis BNR broadcasts live radio shows from 6AM to 7PM. Next to that, BNR also produces podcasts in a wide range of topics. Currently, users can tune-in live to BNR, or listen to historic radio broadcasts, which are offered in their full length on BNR's website, through the BNR mobile app, and through its channels on Spotify and iTunes. Although BNR's archive provides short description of shows, in shows where many different topics are discussed, our users cannot easily tune-in to the parts or sections that are of specific interest to them. Particularly when a user has limited time available, but wants to remain up-to-date on specific topics, this is undesirable. For this purpose with BNR SMART Radio we aim to segment full-length radio shows into shorter coherent pieces of audio. This way, these shorter fragments can be retrieved by users, or served to them through, e.g., personalized playlists.

[3] https://bit.ly/2QLsvQG

Segmenting audio in topically coherent segments is a non-trivial problem. That is why it is more common to provide an interface to search for keywords in an audio stream or some other visualization of topics in the stream [1, 2]. This would, however, be too limited for coherent pieces of radio that can make up a personalized radio show. Since BNR radio is created in-house, we have additional metadata available about the radio shows that we can leverage to inform segmentation of the audio. The most important information that we currently use is the occurrence of jingles that indicate the start and end of segments.

## 3 AUDIO TAGGING

In order to serve segments from the audio segmentation to the right users, each segment needs to be tagged with one or more topics. We automatically transcribe BNR shows using a transcription-service by Zoom Media.[4] We assign tags to segments using a proprietary multilabel text classifier, trained on our own domain-specific dataset, which follows a hand-created taxonomy of tags.

Our current model has a precision of 0.64 and a recall of 0.36. On average the classifier assigns 2.6 tags to each audio segment, with a minimum of 0 and a maximum of 8. Although there is room for improvement, users of the beta version of the BNR SMART Radio app have responded positively to the provided fragments and how it fits their interests, which suggests that the tagging works well in practice.

## 4 SERVERLESS ARCHITECTURE

Managing servers and infrastructure is a time-consuming job and requires special personnel to maintain the system hence costs lots of money, time and headache. By simply leaving the server management to the service provider, we can focus on our building and expanding features. Our goal is to build a system that is reliable, scalable, maintainable and cost-effective.

We achieve this by consuming multiple managed services from Amazon Web Services, e.g., Step Functions and Lambda functions. Step functions are orchestration layer for our functionality in order to transcode, transcribe, and split audio files. These services enable us the visibility of the ongoing operations and scales upon needs effortlessly. See Figure 2 for a more detailed overview of our serverless architecture

## 5 FUTURE WORK

This version of the SMART radio app is only the beginning. It is apparent from the first responses of our testers that creating a personalized list of fragments is appreciated. Therefore we have several ideas to improve the app further.

One area for improvement is the audio tagging. We are aware of limitations of our dataset, for example the size of the sample of tags we used to train our classifier. It might be interesting to look at leveraging a hierarchical structure among the tags. This would have the additional advantage that it would help users browse the taxonomy of tags.

[4]https://www.zoommedia.ai/

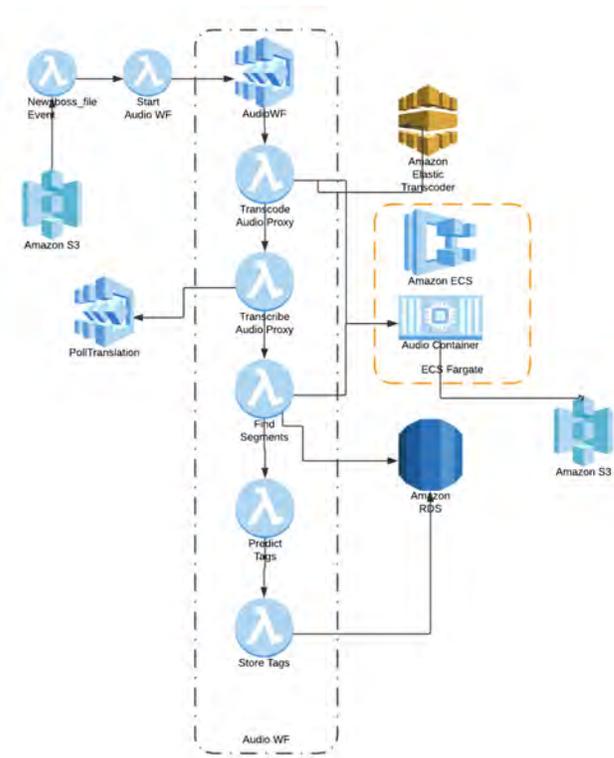

Figure 2: Audio flow

Next, there is space for improvement in the recommendation strategies. The app currently requires users to manually search and select tags they are interested in. In the future we could help users by providing tags suggestions based on their listening behavior, additional user interactions that we have, or relations between tags. Finally, including contextual information such as time, location or device is another direction for improving the recommendations of the app.

Finally, we would like to connect different types of content. Radio shows may provide background or context to news articles, cross-linking FD's news articles with BNR's audio could provide users a more complete news experience, where they can find information from different sources and perspectives more easily.

## 6 ACKNOWLEDGEMENTS

Supported by "Google Digital News Initiative"[5]

## REFERENCES

[1] Marijn Anthonius Henricus Huijbregts. 2008. *Segmentation, diarization and speech transcription: surprise data unraveled.* Citeseer.
[2] Brendan Jou, Hongzhi Li, Joseph G Ellis, Daniel Morozoff-Abegauz, and Shih-Fu Chang. 2013. Structured exploration of who, what, when, and where in heterogeneous multimedia news sources. In *Proceedings of the 21st ACM international conference on Multimedia*. ACM, 357–360.
[3] Maya Sappelli, Dung Manh Chu, Bahadir Cambel, David Graus, and Philippe Bressers. 2018. SMART Journalism: Personalizing, Summarizing, and Recommending Financial Economic News. In *The Algorithmic Personalization and News (APEN18) Workshop at ICWSM '18*.

[5]https://newsinitiative.withgoogle.com/dnifund/

# Position Title Standardization

Rianne Kaptein
Crunchr
Amsterdam, The Netherlands
rianne@crunchrapps.com

## ABSTRACT
In this demo paper we describe a method to standardize position titles of employees. Since position title is typically a free text field in HR (Human Resources) systems, many variations of the same position title can be entered. Reducing the number of different position titles enhances the options for analysis of employee data. We developed a method in which different variations of the same position title are mapped onto one position title taking into account lexical variations, spelling mistakes and synonyms.

## KEYWORDS
Natural Language Processing, Data Cleansing

## 1 INTRODUCTION

A major problem in machine learning and information retrieval alike is data cleansing. Textual data in particular contains a lot of noise. Since it is usually produced by humans textual data is often inconsistent, it contains spelling mistakes and slang and it can be written in different languages or dialects. When designing a system that makes use of this data, the system should either be very robust to these variations in the textual data or a data preprocessing step should be included in the system. In this demo we investigate such a data preprocessing step, namely the problem of position title standardisation.

Position titles are part of the employee data used in our HR analytics software. Position title is one of the factors that can be used to group and filter employees to analyse all kinds of employee data, e.g. the turnover or performance of a group of employees. The issue is that many companies do not use standardized position titles. Position title is usually a free form text field that can be filled out in any language, resulting in thousands of different position titles for a single company. This makes it hard to do any analysis based on this field. In this demo we want to reduce the number of distinct position titles by grouping the position titles that are actually a variation of the same position title.

We have to tackle the following challenges:

- Sometimes small (lexical) differences are essential, i.e. 'CEO' is not the same position title as 'CIO'
- Many abbreviations are used and abbreviations are often company specific. Using a standard list of abbreviations is not enough.

We can exploit however the following characteristics of the data:

- The vocabulary used is limited
- Besides the position title we know also in which location, business unit and functional area this employee works.

For our system it is important that users understand why different position titles are mapped to one position title. Also it should be possible to give feedback on errors the system is making, in particular on company specific information, such as abbreviations.

In the following sections we describe how we adapt to the domain, which transformations are applied, the functionality of our demo, and the results.

## 2 DOMAIN ADAPTATION

Many text processing techniques are available, but it makes quite a difference if you apply them to formal text, social media posts, online information, or in our case position titles. Therefore we take a close look at the applicable text processing techniques and adapt them where needed to make them work for our specific task. We make use of heuristics that we see in our data set to create rules that optimally perform.

We are comparing position titles to each other to try and map multiple position titles onto one (standardized) position title. This means that the order of the transformations of the position titles is very important, e.g. disregarding word order has to be one of the last steps.

For several of our transformations described in the next section, a background vocabulary is needed. From the vocabulary we can use statistical information on how frequent terms occur. Ideally this vocabulary would consist of a very large amount of position titles to create a vocabulary that is similar to the position titles we are analysing. Unfortunately, we do not have such a vocabulary available. Instead, to create our vocabulary we make use of Wikipedia. We use a subset of Wikipedia pages containing pages related to position titles, HR and business information.

Besides the vocabulary, we also adapt the list of abbreviations to the HR domain. We have created a list of common position title related abbreviations, such as jr. for junior and mgr for manager. In addition to this list we try to detect company specific abbreviations by checking against a set of rules. We use the following rules to detect an abbreviation:

- Word consists only of capitals
- Word ends with a dot
- Word is not present in the vocabulary (after spelling correction).

For the detected abbreviations there are three options. The first option is to remove the abbreviation from the position title. This can be done for example for abbreviations that contain information that is also present in other data fields of the employee, such as location or business unit. The second option is to replace the abbreviation with its fully expanded version by adding the abbreviation to a list of company-specific abbreviations. The third option is to do nothing with this abbreviation and just keep it in the position title.

The decision between these options is still a manual step in the process.

## 3 TRANSFORMATIONS

To standardize the position titles, the transformations are applied in the order that follows. For each position title, we keep a reference to its original form.

(1) Capitalization
    All words are lowercased.
(2) Removing terms between brackets
    Terms between brackets are removed from the position title, since they in general contain non-essential information.
(3) Punctuation, including removing hyphens
    All punctuation is removed. For hyphens we check if the words around the hyphen concatenated together form a word from the vocabulary. If so the hyphen is removed and the words are concatenated, otherwise the hyphen is removed and replaced with a space.
(4) Contractions and spaces between words
    We apply the following rules:
    - Double spaces are removed.
    - If a pair of words concatenated together forms a word from the vocabulary or a word from another position title, the pair of words is concatenated, e.g. data base becomes database.
    - Words with a capital in the middle of a word are split, e.g. 'DevOps' is split into 'Dev Ops'.
(5) Abbreviations
    A standard list of abbreviations is used to expand abbreviations, e.g. 'jr.' to 'junior'. In addition to the standard list, there can be company specific abbreviations that are either expanded or removed from the position title.
(6) Spelling mistakes
    To correct spelling we make use of the spelling corrector written by Peter Norvig [2]. For the vocabulary we use the vocabulary we collected from Wikipedia.
(7) Synonyms
    To find synonyms for complete position titles we make use of the redirects in Wikipedia [1]. The Wikipedia redirect structure is highly precise and can be easily extracted from the Wikipedia dumps. Coverage of the position titles is around 9%, i.e. for almost 1 out of 10 position titles we can retrieve one or more synonyms from Wikipedia. For example for the position title 'Chief Human Resources officer' we get the synonyms: 'Vice president in charge of hiring', 'Chief People Officer', 'Chief Personnel Officer', and 'Chief Human Resource Officer'. A larger coverage can be achieved if also partial position titles would be matched, but this increases the probability of introducing errors.
(8) Word order
    Position titles are treated as bags-of-words. Any position title containing the exact same words as another position title is mapped to the same position title.

**Table 1: Similar position titles**

| | |
|---|---|
| Team lead - Customer Service | Customer Service Team Leader |
| Managing Director | CEO |
| QA analyst | Quality Assurance analyst |
| UI/UX Designer | UX/UI Designer |
| Marketer | Marketing Specialist |

## 4 FUNCTIONALITY

The current demo system reads in a set of position titles and then displays to the user three statistics: Possible reduction, number and percentage of distinct position titles. Below this the list of position titles that can be grouped together are listed. The list is grouped by the transformations that have been applied.

## 5 RESULTS

We have applied our position title standardization method on some real datasets, leading to a reduction of around 7% in the number of position titles (e.g. from 2077 to 1934 position titles). The spelling correction makes some errors, e.g 'Backend Engineer' is corrected to 'Backed Engineer'. However, the probability of matching such an incorrectly spelling corrected position title to another position title is very low. The synonyms introduce some mappings that are incorrect or at least debatable, e.g. 'Solutions Architect' vs. 'Technical Architect', and 'Scrum Master' vs. 'Product Owner'. Overall, manual inspection of the results show the precision of our method is very high. Evaluation by the companies HR experts would be needed to present exact numbers on precision. Some examples of position titles that are mapped to the same position titles can be found in Table 1.

Since this functionality can be computed once for a dataset and then reused, performance is not a big issue and we do not report results on it.

As a next step we would like to standardize the position titles further by matching position titles to position titles in a database such as the O*NET database[1]. This should lead to a considerable further reduction in the number of position titles, as well as the opportunity to exploit the structured/hierarchical information on the position titles in the database.

## REFERENCES

[1] P Ipeirotis. Wikisynonyms: Find synonyms using wikipedia redirects. http://wikisynonyms.ipeirotis.com, 2013.
[2] Peter Norvig. How to write a spelling corrector. http://norvig.com/spell-correct.html, 2007.

---

[1]https://www.onetcenter.org/database.html